\documentclass[amsmath,amssymb,aps,prd,superscriptaddress,
reprint
]{revtex4-1}

\usepackage[table,xcdraw]{xcolor}
\usepackage[mathlines]{lineno}
\usepackage{float}
\usepackage{graphicx}
\usepackage{dcolumn}
\usepackage{bm}
\usepackage{hyperref}
\usepackage{siunitx}
\usepackage{graphicx} 
\graphicspath{{Figs/}} 

\usepackage[mathlines]{lineno}
\begin{document}

\title{Dissipative Optomechanics in High-Frequency Nanomechanical Resonators}

\author{Andr\'e G. Primo}
\thanks{These authors contributed equally to this work.}
\affiliation{Gleb Wataghin Institute of Physics, University of Campinas, 13083-859 Campinas, SP, Brazil}

\author{Pedro V. Pinho}
\thanks{These authors contributed equally to this work.}
\affiliation{Gleb Wataghin Institute of Physics, University of Campinas, 13083-859 Campinas, SP, Brazil}

\author{Rodrigo Benevides}
\affiliation{Department of Physics, ETH Zürich, 8093 Zürich, Switzerland}

\author{Simon Gr\"oblacher}
\affiliation{Kavli Institute of Nanoscience, Department of Quantum Nanoscience, Delft University of Technology, Lorentzweg 1, 2628CJ Delft, The Netherlands}

\author{Gustavo S. Wiederhecker}
\affiliation{Gleb Wataghin Institute of Physics, University of Campinas, 13083-859 Campinas, SP, Brazil}

\author{Thiago P. Mayer Alegre}
\email[e-mail:]{alegre@unicamp.br}
\affiliation{Gleb Wataghin Institute of Physics, University of Campinas, 13083-859 Campinas, SP, Brazil}

\date{\today}

\begin{abstract}
\section*{Abstract}
The coherent transduction of information between microwave and optical domains is a fundamental building block for future quantum networks. A promising way to bridge these widely different frequencies is using high-frequency nanomechanical resonators interacting with low-loss optical modes. State-of-the-art optomechanical devices rely on purely dispersive interactions that are enhanced by a large photon population in the cavity. Additionally, one could use dissipative optomechanics, where photons can be scattered directly from a waveguide into a resonator hence increasing the degree of control of the acousto-optic interplay. Hitherto, such dissipative optomechanical interaction was only demonstrated at low mechanical frequencies, precluding prominent applications such as the quantum state transfer between photonic and phononic domains. Here, we show the first dissipative optomechanical system operating in the sideband-resolved regime, where the mechanical frequency is larger than the optical linewidth. Exploring this unprecedented regime, we demonstrate the impact of dissipative optomechanical coupling in reshaping both mechanical and optical spectra. Our figures represent a two-order-of-magnitude leap in the mechanical frequency and a tenfold increase in the dissipative optomechanical coupling rate compared to previous works. Further advances could enable the individual addressing of mechanical modes and help mitigate optical nonlinearities and absorption in optomechanical devices.
\end{abstract}

\maketitle
\section*{Introduction}
The burgeoning field of cavity optomechanics combines the reliability of long-range information transport using optical photons with the versatility of nanomechanical oscillators. This conjunction enabled a plethora of demonstrations, including high-precision force and displacement sensors~\cite{wu_dissipative_2014, liu_integrated_2020, huber_spectral_2020}, and the synchronization of mechanical oscillators~\cite{rodrigues_optomechanical_2021, zhang_synchronization_2012} for signal processing. Furthermore, in high-frequency optomechanical systems, it is possible to actively control the strength of the creation or annihilation scattering process of long-lived phonons~\cite{aspelmeyer_cavity_2014, wiederhecker_brillouin_2019, kippenberg_cavity_2008, he_strong_2020}. This property allows the quantum control of optomechanical systems with promising applications in coherent quantum microwave-to-optical conversion~\cite{jiang_optically_2022, mirhosseini_superconducting_2020, forsch_microwave--optics_2020, jiang_efficient_2020, rueda_efficient_2016, barzanjeh_optomechanics_2022,mckenna_cryogenic_2020, han_microwave-optical_2021, shao_microwave--optical_2019, honl_microwave--optical_2022}, and quantum memories~\cite{wallucks_quantum_2020, stiller_coherently_2020}.

\begin{figure*}[ht!]
\includegraphics[width = 16cm]{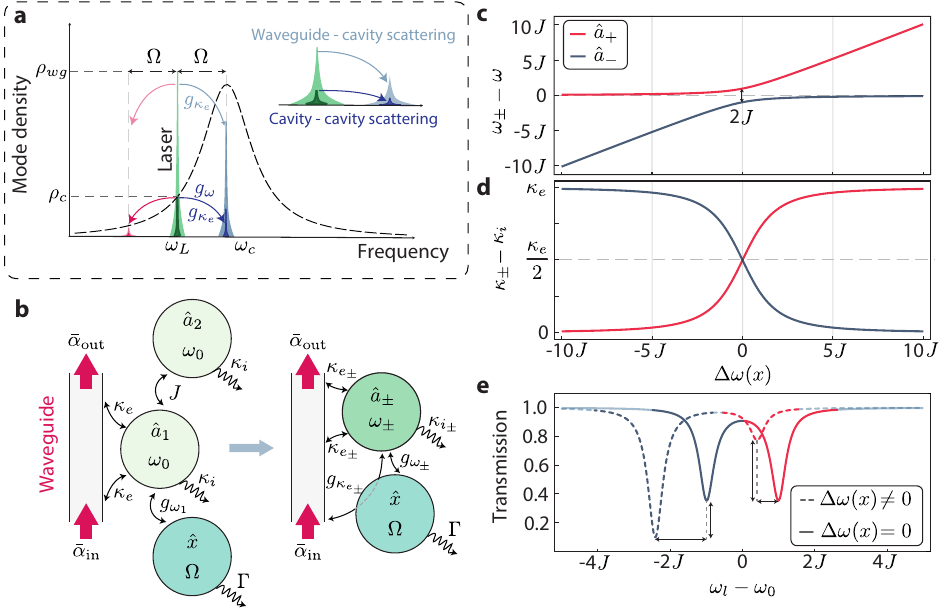}
\caption{\small{\textbf{Concept design for dissipative optomechanical coupled cavity} \textbf{a}, Scattering mechanisms in an optomechanical system with both dispersive and dissipative couplings. Photons arriving through a waveguide (mode density $\rho_\text{wg}$ - light green) can be inelastically scattered into the cavity field depicted by the light blue/red tones. A similar process occurs  for cavity pump photons (mode density $\rho_c$ - dark green), generating sidebands in dark blue/red. \textbf{b}, Diagram for coupled optomechanical cavities with an asymmetric loss induced by the waveguide. The physical response of the system, depicted on the right, necessarily includes a mechanically-dependent extrinsic coupling for both supermodes. \textbf{c}, Frequency splitting in strongly-coupled optical cavities and \textbf{d}, Supermodes' losses as functions of the frequency detuning between the ``bare" resonators, $\Delta\omega(x)$. \textbf{e}, Transmission spectra for different detuning configurations. The trace colors identify the supermodes as in \textbf{c} and \textbf{d}}. Both the extinction and position are sensitive to mechanically coupled detuning, represented by vertical and horizontal double-headed arrows, respectively. }
\label{fig:1}
\end{figure*}

In standard dispersive optomechanical devices, the acoustic modes are engineered to shift the optical cavity resonance frequency, $\omega_0$~\cite{shomroni_optical_2019, chan_laser_2011, fiaschi_optomechanical_2021}. As a consequence, only photons confined to the resonator field are efficiently scattered. This mechanism is quantified by the dispersive optomechanical frequency pulling, $G_\omega = - \frac{d\omega_0}{dx}$, where $x$ is the mechanical displacement amplitude of a given acoustic mode. In dissipatively coupled optomechanical systems, however, the acousto-optic interaction may take place between the optical excitation channel and the cavity~\cite{elste_quantum_2009,weiss_quantum_2013, weiss_strong-coupling_2013, primo_quasinormal-mode_2020}, e.g., through a mechanical modulation of the bus waveguide-cavity coupling rate $\kappa_e$. Its strength is quantified by the dissipative pulling  $G_{\kappa_e} =  \frac{d\kappa_e}{dx}$. In these systems, photons may be scattered directly from the waveguide into the cavity mode, as illustrated in Fig.~\ref{fig:1}\textbf{a}. 

In conjunction with its dispersive counterpart, the presence of the dissipative scattering mechanism leads to remarkable interference phenomena in both mechanical and optical spectra. For instance, the interplay between dispersive and dissipative couplings can vary if a given mechanically-induced shift in the optical frequency is accompanied by either a (mechanically-induced) increase or decrease in optical linewidth, allowing a yet-to-be-explored tool to control optomechanical interactions. Furthermore, using dissipative coupling one could achieve the optomechanical ground-state cooling even in the bad-cavity limit~\cite{elste_quantum_2009, yanay_quantum_2016}, where the mechanical frequency is smaller than the total optical linewidth, $\Omega \ll \kappa$, which is unfeasible using dispersive optomechanics alone. Similar results could be obtained in more complex systems, e.g. using Fano optical lineshapes in hybrid atom-optomechanical systems for sideband suppression/enhancement~\cite{genes_micromechanical_2009}. Although simpler, the dissipative optomechanics approach requires small intrinsic losses, $\Omega  \gg \kappa_i$, a regime thus far elusive~\cite{wu_dissipative_2014, huang_dissipative_2018, li_reactive_2009,barnard_real-time_2019}, hence inhibiting  the application of dissipative optomechanics to coherent information swap protocols~\cite{aspelmeyer_cavity_2014}. State-of-the-art devices still need an order-of-magnitude leap in mechanical frequencies to reach this regime, while preserving low optical losses.

Here, we demonstrate an optomechanical system in the sideband-resolved regime ($\Omega/\kappa \approx 10$) which displays both dissipative and dispersive optomechanical couplings. Our mechanical modes, operating at $\Omega/(2\pi) \approx \SI{5.5}{\GHz}$, represent a two-order-of-magnitude increase in frequency when compared to previous dissipative optomechanics integrated devices~\cite{li_reactive_2009, wu_dissipative_2014}, making our device suitable for applications in the quantum regime such as the optical writing/reading of quantum information into phonons, and optomechanical ground-state cooling. The impact of the mechanically mediated waveguide-cavity interaction was assessed through the acousto-optic transduction in our system, and it displays remarkable signatures on the optomechanical cooling and heating of the mechanical modes, i.e., dynamical backaction~\cite{aspelmeyer_cavity_2014}. In fact, we show that dynamical backaction can be either enhanced or suppressed by controlling the interference between dissipative and dispersive contributions. The effects of dissipative scattering were also visualized, for the first time, on the optical spectrum through the phenomenon of optomechanically induced transparency~\cite{weis_optomechanically_2010,safavi-naeini_electromagnetically_2011}, adding another tool to control integrated tunable optical delays and classical/quantum memories.

\section*{Results}
In Fig.~\ref{fig:1}\textbf{b} we show the principle of operation of our device. Two optical modes, $\hat{a}_1$ and $\hat{a}_2$, with identical frequency $\omega_0$ and intrinsic loss (absorption and radiation) $\kappa_i$, are mutually coupled with a rate $J$. The system is driven through a waveguide, which carries a coherent field with amplitude $\bar{\alpha}_\text{in}$ and couples only to resonator $1$ yielding an extrinsic loss $\kappa_e$. Furthermore, mode $\hat{a}_1$ is dispersively coupled to a mechanical mode $\hat{x}$ with a frequency pulling parameter $G_{\omega_1}$, inducing a detuning $\Delta\omega(x) = -G_{\omega_1} \hat{x}$ between $\hat{a}_1$ and $\hat{a}_2$. This detuning changes the effective coupling between resonators  $\hat{a}_1$ and $\hat{a}_2$, giving rise to a mechanically-dependent collective response of the system, described by the supermodes $\hat{a}_+$ and $\hat{a}_-$~\cite{burgwal_enhanced_2023, jayich_dispersive_2008, miri_exceptional_2019}. 

\begin{figure*}[ht!]
\centering
\includegraphics[width = 16cm]{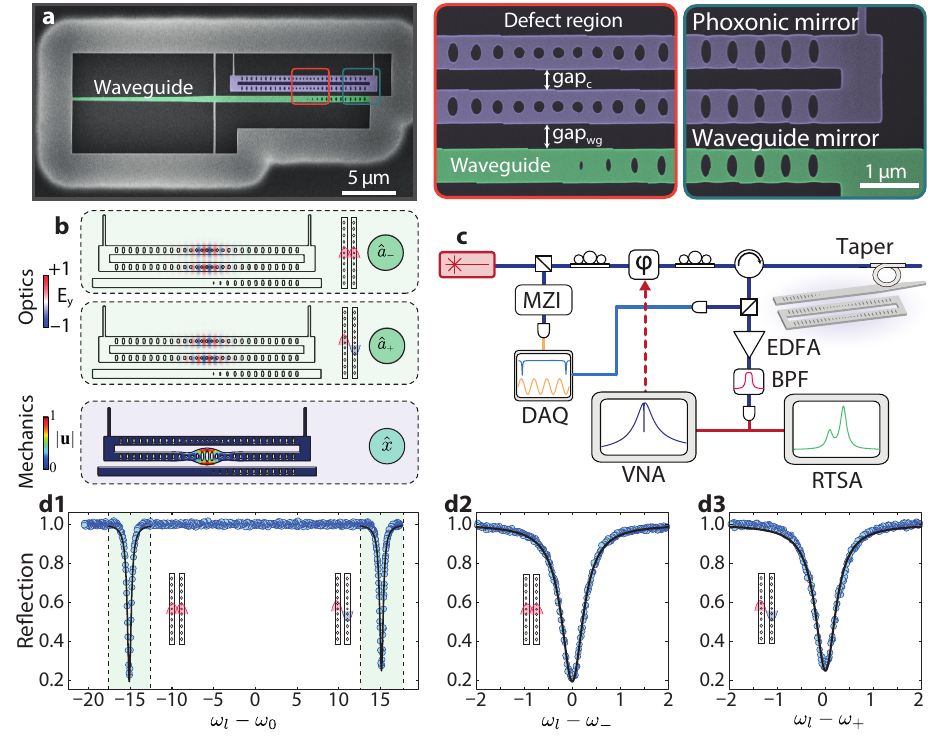}
\caption{\small{\textbf{Experimental setup and characterization of coupled optomechanical resonator} \textbf{a}, Scanning electron micrograph of one of the fabricated devices. The image is false-colored to highlight the coupled optomechanical resonators (blue) and tapered waveguide (green). The defect region is responsible for the high-quality confinement of both mechanical and optical modes. The phoxonic (photonic and phononic) mirror suppresses any mechanical coupling between the two nanobeams. \textbf{b}, Finite-element Method simulations of the optical ($\hat{a}_-$, $\hat{a}_+$) and mechanical breathing ($\hat{x}$) modes of our system. Here, we plot the normalized $y$-component of the electric field, $E_y$, and mechanical displacement $|\vec{u}|$. Although both beams support identical mechanical modes, only one is shown for simplicity. \textbf{c}, Schematics of the measurement setup. A tunable laser drives our device, which is accessed using a tapered fiber coupled to the integrated waveguide. The thermo-mechanical noise imprinted in the reflected light is collected and characterized with a real-time spectrum analyzer (RTSA). Coherent spectroscopy is performed using a vector network analyzer (VNA), which modulates the phase of the driving field with an electro-optic phase-modulator ($\phi$). Reflection spectra are measured with a DAQ, along with the output of a Mach-Zehnder interferometer (MZI), which provides the relative frequency of our laser. \textbf{d1}, Normalized reflection spectrum of the cavity, showing common and differential optical modes. The region around the two resonances is finely scanned with a laser and fitted to a Lorentzian model, as shown in  \textbf{d2} and \textbf{d3}. For the device under analysis, $\text{gap}_\text{c} = \SI{500}{\nano\metre}$ and $\text{gap}_\text{wg} = \SI{450}{\nano\metre}$.}}
\label{fig:2}
\end{figure*}

In general, the supermodes' frequencies ($\omega_\pm$ - Fig.~\ref{fig:1}\textbf{c}) and losses ($\kappa_\pm$ - Fig.~\ref{fig:1}\textbf{d}) will depend on the position of the mechanical oscillator, as their response is a combination of the individual properties of $\hat{a}_1$ and $\hat{a}_2$. In fact, in the absence of the mechanically-induced detuning, $\Delta\omega(x) = 0$, the supermodes are simply the differential, $\hat{a}_+ =  (\hat{a}_1-\hat{a}_2)/\sqrt{2}$, and common, $\hat{a}_- =  (\hat{a}_1 +\hat{a}_2)/\sqrt{2}$, mode pairs. As such, electromagnetic energy is evenly divided between $\hat{a}_1$ and $\hat{a}_2$. Since the mechanical motion drives the system away from this regime, it leads to an asymmetric optical field distribution among the resonators, and the supermode with larger energy density overlap with cavity $1$ will have larger extrinsic losses, due to its coupling to the bus waveguide. This interplay leads to a dissipative coupling~\cite{yanay_quantum_2016} ($G_{\kappa_{e_\pm}} = \frac{d\kappa_{e_\pm}}{dx}$) in addition to the usual dispersive coupling $G_{\omega_\pm} = \frac{d\omega_\pm}{dx}$. The joint action of these effects is illustrated in  Fig.~\ref{fig:1}\textbf{e}, where $\Delta\omega(x)$ leads to variations in both frequencies and extinctions (losses) of the  $\hat{a}_\pm$ supermodes.

From our analysis, both the effective dispersive and dissipative couplings are dependent on the individual dispersive coupling of cavity 1, $G_{\omega_1}$. Furthermore, when a weaker intercavity coupling $J$ is present, it results in a steeper avoided crossing between the bare optical modes $a_1$ and $a_2$. This characteristic makes the system more susceptible to mechanical perturbations, thereby increasing the dissipative couplings. However, it is important to ensure that the adiabaticity condition $2J \gg \Omega$ is satisfied (refer to section~\textbf{S}1 of the Supplemental Material for a detailed mathematical discussion).

We implement this scheme using a pair of identically-designed silicon photonic crystal nanobeams, as shown in Fig.~\ref{fig:2}\textbf{a} (see Methods). An engineered defect in the central region of both nanobeams supports co-localized optical and mechanical modes with resonances in the optical ($\SI{1550}{nm}$) and microwave ($\Omega/(2\pi) \approx \SI{5.5}{GHz}$) bands, respectively. The resonators are separated by a gap, $\text{gap}_c$, which evanescently couples their optical modes, in contrast to their acoustic modes which are uncoupled due to an efficient phononic mirror at the clamping edges of the nanobeams. A waveguide is placed laterally to one of the optomechanical cavities and used to probe the device. A photonic mirror defined in the bus waveguide ensures that the output field is efficiently collected. Finite element method simulations for the electromagnetic and acoustic modes of the system are shown in Fig.~\ref{fig:2}\textbf{b}. The devices were characterized using the setup shown in Fig.~\ref{fig:2}\textbf{c}, which can probe both optical and mechanical properties of the device at room temperature. The optical response is characterized by scanning the frequency of a continuous-wave tunable laser and monitoring the cavity's reflection spectrum (see Methods). The coupled optical modes of our system appear as two sharp resonances as shown in Figs.~\ref{fig:2}\textbf{d}.

\begin{figure*}[ht!]
\includegraphics[width = 16cm]{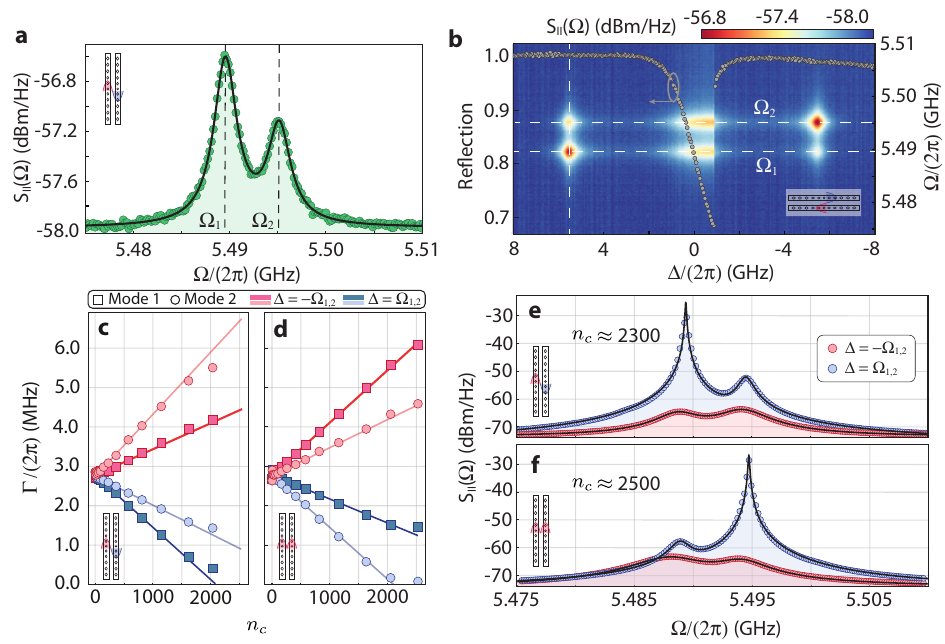}
\caption{\small{\textbf{Dissipative and dispersive contributions to optomechanical backaction} \textbf{a}, Thermo-mechanical spectrum of the differential optical mode in our device at an input power $P_\text{in} = \SI{2.7}{\micro\watt}$. Each peak corresponds to the acoustic breathing mode of individual nanobeams. \textbf{b}, Map of the mechanical spectra as a function of the laser-cavity detuning $\Delta$. The vertical dashed line corresponds to the spectrum in \textbf{a}. The reflection at every $\Delta$ in our measurement is also provided; the gray arrow indicates the associated axis. \textbf{c} (\textbf{d}) Optically-induced modifications to the mechanical linewidth of mechanical modes 1 and 2 for the differential (common) optical modes, as a function of the number of photons in the resonator. \textbf{e}, Selective phonon lasing contrasted to optomechanical cooling in the differential optical mode. \textbf{f}, The same analysis for the common optical mode.}}
\label{fig:3}
\end{figure*}

The optomechanical interaction imprints information about the thermomechanical motion of the cavity onto the reflected light. Fig.~\ref{fig:3}\textbf{a} shows the optical detector's photocurrent power spectral density, $S_\text{II}(\Omega)$. Two acoustic modes were found around $\SI{5.5}{GHz}$, corresponding to the breathing modes of each silicon nanobeam. Due to natural variations in the fabrication process (see Methods), these modes are non-degenerate and have frequencies $\Omega_1$ and $\Omega_2$. Lorentzian fittings yield mechanical Q-factors $Q_m \approx 2100$ for both modes. Fig.~\ref{fig:3}\textbf{b} shows a density map constructed by stacking multiple photocurrent spectral densities obtained for a range of laser-cavity detunings around the differential optical supermode, $\Delta = \omega_l - \omega_+$; the data in Fig.~\ref{fig:3}\textbf{a} is represented by the vertical dashed line. The optical mode reflection is also plotted for reference as overlaid points in this map showing a slightly thermo-optical induced bistable regime.

Analyzing $S_\text{II}(\Omega)$ in Fig.~\ref{fig:3}\textbf{b} we see clear mechanical signals at $\Delta = \pm \Omega_{1,2}$. Here, input powers
are kept low to avoid linewidth modifications in the acoustic spectrum (intracavity photon occupation of $n_c \approx 14$). Considering the response of mechanical modes $1$ and $2$ as $\Delta$ is varied, we verify a clear imbalance between their signals at $\Delta =  \Omega_{1,2}$ and $\Delta = -\Omega_{1,2}$ \color{black}(see an example of this imbalanced response as a function of the optical detuning in section \textbf{S6}, and FIG.~\textbf{S1} of the Supplemental Material)\color{black}. In the sideband-resolved regime, $\Delta =  \pm \Omega_{1,2}$ is the laser-cavity detuning of interest since either the Stokes or Anti-Stokes sidebands generated by the optomechanical interaction are resonantly enhanced, thus leading to appreciable backaction heating or cooling at large input powers.  The imbalance in our data suggests that the rates for Stokes and Anti-Stokes scatterings are different for each of the mechanical modes, although with opposing trends for modes $1$ and $2$ -- while at $\Delta =  \Omega_1$ the signal for mechanical mode $1$ is enhanced, it is suppressed for $2$. These results, showing a novel degree of control enabled by the optomechanical interaction, cannot be explained solely within the framework of dispersive optomechanics where the probabilities for Stokes and Anti-Stokes scatterings are identical when considering a system far from ground-state, as is the present case 

Our measurements are compatible with a system presenting both dispersive and dissipative optomechanical couplings~\cite{wu_dissipative_2014, huang_dissipative_2018}. In this scenario, these two scattering mechanisms can interfere constructively or destructively, depending on the laser-cavity detuning $\Delta$, leading to different scattering rates for Anti-Stokes and Stokes bands. Yet, such behavior has not been previously observed in the resolved sideband regime. The root of the observed asymmetry between signals at $\Delta =  \pm \Omega_{1,2}$ lies in the different phases of the intracavity and waveguide fields, which enhance dispersive and dissipative scattering mechanisms, respectively. In a frame rotating at the laser frequency, the phase of the cavity field is shifted by $\approx \pi$ between red and blue detunings whereas the waveguide field has a fixed phase. Consequently, the two mechanisms' interference varies from constructive to destructive between blue and red detunings. Furthermore, the opposite behavior of mechanical modes $1$ and $2$ can be explained through an extension of the setup of Fig.~\ref{fig:1}\textbf{b}: the mechanically induced detuning $\Delta\omega(x)$ has a different sign if the acoustic mode is coupled to optical mode $2$ rather than $1$. Since the variation in the optical linewidth is proportional to $\Delta\omega(x)$, mechanical modes $1$ and $2$ will have dissipative couplings with opposite signs (for a given optical supermode), which in turn results in dissipatively scattered fields with opposite phases.

At sufficiently high optical input powers, driving our system at frequencies resonant with Stokes ($\Delta = \Omega_{1,2}$) or Anti-stokes ($\Delta = -\Omega_{1,2}$) scattering processes leads to appreciable amplification or cooling of the mechanical modes. We anticipate an imbalance in the efficiency of these processes inherited by the same interference effects
between the dispersive and dissipative scattering channels discussed above. In the sideband-resolved limit the extra damping rate of the mechanical mode, $\Gamma^\text{OM} (\Delta)$, is given by (see~\textbf{S}3)
\begin{equation}
    \Gamma^\text{OM} (\pm \Omega) \approx \mp  \frac{ ( \mp \, 2\sqrt{n_c} g_\omega + g_{\kappa_e} \frac{\bar{\alpha}_\text{in}}{\sqrt \kappa_e})^2}{\kappa},
    \label{eq:1}
\end{equation}
where $g_\omega = G_\omega x_\text{zpf}$ and $g_{\kappa_e} = G_{\kappa_e} x_\text{zpf}$ are the dispersive and dissipative vacuum optomechanical coupling rates of (any of) the supermodes, and $x_\text{zpf}$ is the zero-point fluctuation in the mechanical displacement. From Eq.~\ref{eq:1}, we expect different relative signs of $g_{\kappa_{e_\pm}}/g_{\omega_{\pm}}$ to benefit opposite backaction effects (cooling or heating). For instance, if $g_\kappa/g_\omega < 0$ the heating process is enhanced, to the detriment of cooling, whereas the converse happens if $g_\kappa/g_\omega > 0$. An interesting feature of Eq.~\ref{eq:1} is the dissipative contribution scaling with the normalized input field amplitude $\bar{\alpha}_\text{in}/\sqrt{\kappa_e}$, which is a factor $\Omega/\kappa_e$ larger than $\sqrt{n_c}$ in the sideband-resolved regime, \color{black} thus leading to an appreciable dissipative contribution to optomechanical backaction even in the regime of $|g_{\kappa_{e_\pm}}/g_{\omega_{\pm}}|\ll 1$. 
\color{black}
 
The optical control of the mechanical linewidth in our device is shown in Fig.~\ref{fig:3}\textbf{c} for the differential optical mode. Due to dissipative coupling and its opposite signs between acoustic modes, one mechanical mode can be preferentially heated or cooled over the other. When driving at the common optical mode the responses of the mechanical modes are swapped, as shown in Fig.~\ref{fig:3}\textbf{d}. This is once more consistent with the analysis presented in Fig.~\ref{fig:1}\textbf{d}: the variation in the losses of each of the supermodes $\kappa_\pm$ is opposite for given a detuning $\Delta\omega(x)$, hence yielding dissipative couplings with different signs for $a_+$ and $a_-$. The solid lines represent fits of the full model of Eq.~\ref{eq:1} to the experimental data from which we extract both dispersive and dissipative coupling rates between all the mechanical and optical modes. The obtained values are displayed in Table~\ref{tab:1}. For sufficiently large photon occupations $n_c>2000$  we achieve photon-phonon cooperativities $C =  |\Gamma^\text{OM}|/\Gamma > 1$. This is an important metric for quantum and classical information transfer protocols whose efficiency scales with $C$. Such large cooperativities are evidenced by the mechanical spectra in the differential (Fig.~\ref{fig:3}\textbf{e}) and common (Fig.~\ref{fig:3}\textbf{f}) optical modes under blue (red) laser-cavity detunings where clear narrowing (broadening) of the mechanical modes is verified. Interestingly, the presence of the dissipative coupling and its different signals between mechanical modes allows one to selectively induce self-sustained oscillations in each of them.

\begin{table}[h!]
\centering
\begin{tabular}{c|cc|cc|}
\cline{2-5}
 & \multicolumn{2}{c|}{\textbf{Mech. Mode 1}} & \multicolumn{2}{c|}{\textbf{Mech. Mode 2}}\\\cline{2-5}  
 & \multicolumn{1}{c|}{$g_{\omega}/2\pi \; (\SI{}{kHz})$} & $g_{\kappa}/2\pi \; (\SI{}{kHz})$ & \multicolumn{1}{c|}{$g_{\omega}/2\pi \; (\SI{}{kHz})$} & $g_{\kappa}/2\pi \; (\SI{}{kHz})$ \\ \hline
\multicolumn{1}{|c|}{$a_+$} & \multicolumn{1}{c|}{$352\pm 10$} & $-2.8\pm 0.1$ & \multicolumn{1}{c|}{$378\pm 10$} & $3.4\pm 0.1$ \\ \hline
\multicolumn{1}{|c|}{$a_-$} & \multicolumn{1}{c|}{$356\pm 10$} & $3.8\pm 0.1$ & \multicolumn{1}{c|}{$367\pm 10$} & $-3.1\pm 0.1$ \\ \hline
\end{tabular}
\caption{Dissipative and dispersive coupling rates for each pair of optical and mechanical modes extracted from fits of the full model leading to Eq.~\ref{eq:1}. }
\label{tab:1}
\end{table}

\begin{figure*}[ht!]
\centering
\includegraphics[width = 16cm]{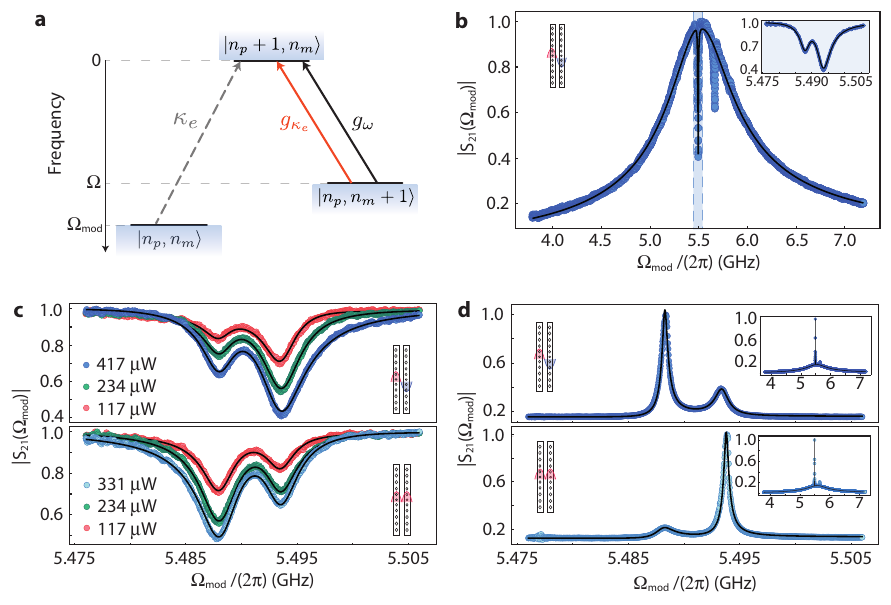}
\caption{\small{\textbf{Optomechanical induced transparency from dissipative and dispersive coupling} \textbf{a}, Level diagram of OMIT. The probe beam induces phonon-conserving transitions while the mechanically-scattered photons necessarily decrease the phonon population. The two optomechanical coupling mechanisms (dissipative and dispersive) interfere to generate the transparency window. \textbf{b}, Scattering parameter magnitude, $|S_{21}|$, as a function of the phase-modulation frequency, $\Omega_\text{mod}$. Data is shown for the differential optical mode. Inset: Transparency window as highlighted in blue. \textbf{c}, Transparency windows for the differential (top) and common (bottom) optical modes as a function of input power. \textbf{d}, Absorption windows for the differential (top) and common (bottom) optical modes. Input powers are indicated through the matching colors with \textbf{c}.}}
\label{fig:4}
\end{figure*}

In the sideband-resolved regime photons and phonons can hybridize and give rise to photon-phonon polaritons~\cite{safavi-naeini_electromagnetically_2011, weis_optomechanically_2010, weiss_strong-coupling_2013}, marking the onset of the strong-coupling regime of optomechanics. This phenomenon is more noticeable if the system is driven resonantly with the mechanical mode, i.e. $\Delta = \pm \Omega$. In this case, transmission and reflection measurements clearly display features arising from the mechanical lineshape, appearing as either dips or peaks in the optical spectrum and leads to substantial modifications in the group velocity of light passing through the device. This effect is the optomechanical analog of electromagnetically-induced transparency/absorption in atomic~\cite{zhang_creation_2009} and solid-state systems~\cite{phillips_electromagnetically_2003} and hence named optomechanically-induced transparency or absorption (OMIT/OMIA).

Regardless of the recent progress in dissipative optomechanical systems, the observation OMIT/OMIA remains elusive since the resolved sideband regime has not yet been reached~\cite{li_reactive_2009,wu_dissipative_2014,huang_dissipative_2018}. Our device allies large mechanical frequencies and dissipative couplings, making it uniquely suited for such demonstration. OMIT is achieved by setting a carrier laser, at frequency $\omega_l$, such that its mechanically scattered Anti-Stokes sideband is resonant with an optical mode, i.e., $\Delta = -\Omega$. Fundamentally, the optomechanical interaction generated by the pump causes transitions annihilating one phonon ($n_m \rightarrow n_m-1$), where $n_m$ denotes the phonon occupation. On the optical domain, energy conservation requires a frequency-shifted photon (population $n_p$) to be created ($n_p \rightarrow n_p+1$). A probe beam, at frequency $\omega_l+ \Omega_\text{mod}$, induces phonon-number conserving transitions. When $\Omega_\text{mod}$ approaches $\Omega$, interference between mechanically scattered photons and the probe beam results in a transparency window in the probe's reflection spectrum. In our system, the depth and width of the dip in the optical spectrum are largely affected by both dispersive and dissipative mechanical scattering mechanisms, leading to the first observation of dissipative signatures in OMIT/OMIA. A generalized scheme for OMIT, accounting for dissipative and dispersive effects, is summarized in Fig~\ref{fig:4}\textbf{a}.

We assess the changes in the reflected optical spectra by phase-modulating our strong input laser (pump) and generating a weak probe beam. The modulation frequency, $\Omega_\text{mod}$, is varied using a vector network analyzer (VNA), which also measures the beating signal between the pump and the probe beams, yielding the scattering parameter $S_{21}(\Omega_\text{mod})$ (see Methods). A typical curve for $|S_{21}(\Omega_\text{mod})|$ is shown in Fig~\ref{fig:4}\textbf{b}, where we highlight the aforementioned transparency window, at a frequency matching the nanobeams' breathing modes.

In Fig.~\ref{fig:4}\textbf{c} we zoom into the transparency windows measured for both optical modes for a range of input powers and detunings $\Delta \approx  - \Omega_{1,2}$. Remarkably, we selectively induce deeper and wider dips at $\Omega_1$ ($\Omega_2$) by driving the common (differential) optical mode. We repeat this experiment for a range of input powers and detunings $\Delta \approx  - \Omega_{1,2}$. A model for $|S_{21}(\Omega_\text{mod})|$ including optomechanical effects is fit to the data and yields $g_{\kappa_{e}}$  and $g_{\omega}$ consistent with our previous characterization (see Methods). Our experiments are limited by the thermo-refractive response of the cavity, which inhibits access to $\Delta  = -\Omega_{1,2}$ at large input powers. This could be mitigated by cooling the present device to temperatures $T<\SI{20}{\kelvin}$, where the mechanical Q-factors are increased and optically-induced heating is reduced. In Fig.~\ref{fig:4}\textbf{d} we display complementary OMIA measurements, where the pump is set to $\Delta \approx \Omega_{1,2}$, and once more the mechanical mode selectivity is observed. Our theoretical model, fed with the parameters extracted from the OMIT experiments, was able to reproduce the OMIA results with great accuracy, further validating our analysis.

We demonstrated the first dissipative optomechanical system operating in the good-cavity regime. To the best of our knowledge, our results represent a two-order of-magnitude leap in the acoustic frequencies ($\SI{5.5}{\giga\hertz}$)~\cite{li_reactive_2009} with a tenfold increase in dissipative couplings ($|G_{\kappa_{e_\pm}}| \approx \SI{1}{\giga\hertz/\nano\metre}$)~\cite{huang_dissipative_2018} when compared to previous literature. This allowed the first observations of dissipative optomechanics signatures on both optical and mechanical spectra, \color{black} despite the modest figure of $|g_{\kappa_{e}}/g_{\omega}|\approx 1\%$. Our data and theoretical modeling showed that the waveguide field enhancement in the dissipative optomechanical interaction (in contrast to the intracavity field enhancement in dispersive optomechanics) boosts the dissipative contribution in the interplay between acoustic and optical responses in our device, mitigating the shortcomings of small $g_{\kappa_e}$.
\color{black}
In our system, an approximate sixfold enhancement in $g_{\kappa_e}$ could be achieved solely by reducing the photon tunneling rate, enabling, for example, an almost complete cancellation of either the backaction heating or cooling in a given optical/mechanical mode pair. In this condition, one obtains almost independent optomechanical control of the virtually degenerate mechanical modes $1$ and $2$ \color{black} (see section \textbf{S7}, and FIG.~\textbf{S2} of the Supplemental Material).\color{black} Individual addressing of acoustic modes could also be accomplished in dual-tone experiments with simultaneous red- and blue-detuned lasers. By tuning the input power of each tone one could completely cancel backaction on a given mechanical mode. In the classical realm, this prospect is potentially interesting for sensing applications, as in the generation of mechanical exceptional points~\cite{wu_-chip_2023}, where tuning losses of different mechanical modes is necessary. In the quantum regime, artificial atoms such as microwave qubits and color centers could be individually addressed by each mechanical mode, creating a scalable tool for the accurate control of quantum information~\cite{prabhu_individually_2023, shandilya_optomechanical_2021}, moreover, this setup could be further explored for quantum sensing, e.g., in backaction-evading experiments~\cite{shomroni_optical_2019}. 

A critical analysis of Eq.~\ref{eq:1} reveals that dissipative scattering could overcome its dispersive counterpart if $g_\omega < g_{\kappa_e} \Omega/(2\kappa_e)$. State-of-the-art dispersive systems currently operate at $g_\omega/(2\pi) \approx \SI{1}{\MHz}$, thus, the predominance of the dissipative coupling requires $g_{\kappa_e} > \SI{54}{\kHz}$ (for the mechanical frequencies and extrinsic losses of our device), which is over one order of magnitude larger than the values reported here. However, as shown in our data, dispersive and dissipative contributions can constructively interfere and therefore having nonzero $g_{\kappa_e}$ can be advantageous even at smaller values. Complementarily, a plethora of design routes~\cite{dumont_flexure-tuned_2019, tagantsev_dissipative_2021, xuereb_dissipative_2011} (with integrated photonic analogs) and toolboxes~\cite{primo_quasinormal-mode_2020} for dissipative optomechanics can be used to further improve our figures. Advances in this field could also be obtained by exploring more complex systems such as cavity exciton-polaritons~\cite{kyriienko_optomechanics_2014}, and electromechanical resonators~\cite{elste_quantum_2009}.
 
As a last remark, the waveguide-cavity scattering process discussed and demonstrated in this work remains vastly unexplored. It enables the optomechanical interaction to take place even in the absence of circulating pump photons in the cavity, which is potentially an advantage over the dispersive process. One could harness this advantage by introducing devices that entirely suppress the intracavity field, as in a system interacting with two driving channels destructively interfering at the cavity. In this case, the dissipative coupling allied to strong excitation fields could still generate mechanically scattered photons in the optical mode, while the resonator's optical response would be cloaked against unwanted linear/nonlinear absorption arising from the buildup of pump photons in the cavity mode. This is currently one of our limitations in achieving high optomechanical cooperativities.


\section*{Methods}
\subsection{Fabrication}

The device manufacturing procedure follows a basic CMOS-compatible top-down approach following the recipe developed for Ref.~\cite{fiaschi_optomechanical_2021}. Electroresist polymer (CSAR-09) is spun at 2000 rpm for 1 min on top of an SOI (silicon on insulator) wafer die. The device designs are patterned on the electroresist using a $\SI{100}{kV}$ e-beam lithography tool followed by a development step immersing the chip for 1 minute in the pentyl-acetate solution. The device is then transferred to the silicon layer using a SF$_6$+O$_2$ plasma etch at cryogenic temperatures. The residual electroresist is removed using a piranha process (H$_2$SO$_4$:H$_2$O$_2$ - 3:1) followed by silica etch to release the structure using HF (hydrofluoric acid) solution for 3 min. 

We attribute to fabrication variations the observed shift in the mechanical frequencies of resonators 1 and 2 to mainly: \textbf{a)}  global electron dosing variations due to the writing pattern, \textbf{b)} turbulence during the insertion of the chip into the chemical solution for electron resist development, which can produce variations in the development rate, and \textbf{c)} temperature gradient in the chip during plasma etching influencing the local etch rate. Further details on the reproducibility of our devices and variations in their critical properties can be found in the Supplementary Material of Ref.~\cite{fiaschi_optomechanical_2021}.

\subsection{Optical Characterization}
The optical characterization was performed using a Toptica CTL 1550 laser. From the data shown in Fig.~\ref{fig:2} we infer a photon tunneling rate $J/2\pi \approx \SI{15}{GHz}$ from the mode-splitting. A Lorentzian model fit to each transmission dip extracts their linewidths, $(\kappa_{-},\,\kappa_{+})/2\pi = (546,\, 515)$~MHz, and extrinsic losses $(\kappa_{e_-},\,\kappa_{e_+})/2\pi = (153,\,129)$~MHz.

\subsection{Optomechanical Transduction}
 The reflected light was captured using a fast photodetector (Discovery Semiconductors DSC30S) and a real-time spectrum analyzer (Agilent PXA N9030A). The laser frequency is swept from red to blue detunings, allowing the access of $S_\text{II}(\Omega)$ at $\Delta = \pm \Omega_1$, which becomes challenging for high input powers due to optical nonlinearities leading to a bistable behavior on the optical spectrum~\cite{primo_accurate_2021,barclay_nonlinear_2005}. In our experiment, optical bistability is observed for optical powers as low as $\SI{1}{\micro\watt}$. Transduction with optical power below the bistable regime is shown in~\textbf{S}6.
 
 The results are modeled using
 input-output theory to describe the photocurrent power spectral density $S_{II}(\Omega)$ shown in Fig.~\ref{fig:3}. The classical optical field amplitude, $a(t)$, is given by

\begin{equation}
\dot a(t) = i (\Delta + G_\omega x )a(t)  -\frac{\kappa + G_{\kappa_e}x }{2}a(t) -\sqrt{\kappa_e} \bar{\alpha}_\text{in} - \frac{G_{\kappa_e}x}{2\sqrt{\kappa_e}}  \bar{\alpha}_\text{in}.
\label{eq:rate_a}
\end{equation}
Linearizing this equation around stationary coherent amplitudes $\bar{x}$ and $\bar{a}$, i.e., $a(t) \to \bar{a} + \delta a (t)$, $x(t) \to \bar{x} +  \delta x (t)$, and keeping terms only to first order in the fluctuations $\delta x (t)$, $\delta a (t)$, we arrive at the output field amplitude

 \begin{equation}
     \alpha_\text{out}(t) =  \bar{\alpha}_\text{in} + \sqrt{\bar{\kappa}_e} \left(\bar{a}+\delta a(t)\right)+\frac{G_{\kappa_e}\bar{a}}{2\sqrt{\bar{\kappa}_e}} \delta x (t).
     \label{eq:outp}
 \end{equation}

The photocurrent $I(t)$ is proportional to $|\alpha_\text{out}(t)|^2$, whose fluctuations are given by $\delta I(t) = \bar{\alpha}_\text{out} \delta \alpha^*_\text{out}(t)+\bar{\alpha}^*_\text{out} \delta \alpha_\text{out}(t)$. Using a spectrum analyzer, one measures the power spectral density of $\delta I(t)$, $S_{II}(\omega) = \int d \tau e^{i \omega \tau}\langle \delta I(\tau) \delta I(0)\rangle$. Finding this quantity requires moving into a frequency domain description in Eq.~\ref{eq:outp}, where the optical field $\delta a(\omega)$ is written in terms of $\delta x(\omega)$. Finally,  $S_{II}(\omega)$ can be written in terms of the mechanical power spectral density, $S_{xx}(\omega)$, which is independent of the detuning for low input powers. The transduction function between $S_{xx}(\omega)$ and $S_{II}(\omega)$ is given in \textbf{S}2 and depends on the detuning $\Delta$, $\omega$ and other properties such as the extrinsic coupling and total losses of the optical mode under analysis.

\subsection{Optomechanically Induced Transparency}

The phase-modulation on the pump frequency, $\Omega_\text{mod}$, is varied using a vector network analyzer (Agilent PNA E8362C). The carrier and probe are reflected from the cavity and interfere at a fast photodiode, yielding a fluctuating photocurrent that carries the information of any optomechanical contribution to the probe's spectrum. This photocurrent is fed to the VNA which measures the scattering parameter $S_{21}(\Omega_\text{mod})$.
 
The input laser phase modulation is described as  $\bar{\alpha}_\text{in} \to \bar{\alpha}_\text{in}e^{-i \phi_0 \sin(\Omega_\text{mod}t)}$ in Eq.~\ref{eq:rate_a}. For weak modulations, $\phi_0 \ll 1$, the system is effectively driven by a strong pump tone, at frequency $\omega_l$ and two probes at $\omega_l \pm \Omega_\text{mod}$. In the sideband-resolved regime, $\Omega_\text{mod} \gg \kappa$ and the cavity response filters out one of the probe tones.  The remaining sideband induces a cavity amplitude $a_+$, which is given by

\begin{equation}
    a_+ \approx -\frac{\sqrt{{\kappa}_e}\bar{\alpha}_\text{in}\phi_0}{-2i\left( {\Delta} + \Omega_\text{mod}\right) + {\kappa} + \frac{n_c \left({\Delta} g_{\kappa _e}-2 {\kappa}_e g_{\omega
   }\right)^2}{{\kappa}_e^2 \left(\Gamma -2 i \Omega
   _{\text{mod}}+2 i \Omega \right)}},
   \label{eq:a+}
\end{equation}
where we assumed $\Delta < 0$ and $|\Delta| \gg  \kappa$, leading to the OMIT configuration. Here $\Omega$ is the mechanical frequency of the acoustic mode under analysis and $\Delta$, $\kappa_e$, and $\kappa$ already include static shifts due to an average mechanical displacement $\bar{x}$.

We immediately verify that the optical susceptibility is strongly dressed by the optomechanical interaction when $\Omega_\text{mod}\approx \Omega$. This information is naturally imprinted in the reflection spectrum of the probe, which is directly connected to the magnitude of the scattering parameter $S_{21}(\Omega_\text{mod})$, measured in our experiment. A full derivation of this equation is presented in \textbf{S}4.

An extension of Eq.~\ref{eq:a+} can be used to derive a model for $|S_{21}(\Omega_\text{mod})|$ (see \textbf{S}4).  Fixing the ratio $g_{\kappa_{e}}/g_\omega$ using the results from Fig~\ref{fig:3} and fitting $|S_{21}(\Omega_\text{mod})|$ to the data in Fig.~\ref{fig:4} we obtain the results in Table~\ref{tab:2}.
 
\begin{table}[h!]
\centering
\begin{tabular}{c|cc|cc|}
\cline{2-5}
 & \multicolumn{2}{c|}{\textbf{Mech. Mode 1}} & \multicolumn{2}{c|}{\textbf{Mech. Mode 2}}\\\cline{2-5}  
 & \multicolumn{1}{c|}{$g_{\omega}/2\pi \; (\SI{}{kHz})$} & $g_{\kappa}/2\pi \; (\SI{}{kHz})$ & \multicolumn{1}{c|}{$g_{\omega}/2\pi \; (\SI{}{kHz})$} & $g_{\kappa}/2\pi \; (\SI{}{kHz})$ \\ \hline
\multicolumn{1}{|c|}{$a_+$} & \multicolumn{1}{c|}{$349\pm 8$} & $-2.8\pm 0.1$ & \multicolumn{1}{c|}{$405\pm 10$} & $3.6\pm 0.1$ \\ \hline
\multicolumn{1}{|c|}{$a_-$} & \multicolumn{1}{c|}{$373\pm 10$} & $4.0\pm 0.1$ & \multicolumn{1}{c|}{$372\pm 10$} & $-3.1\pm 0.1$ \\ \hline
\end{tabular}
\caption{Dissipative and dispersive coupling rates for each pair of optical and mechanical modes extracted from OMIT measurements.}
\label{tab:2}
\end{table}

\section*{Data Availability}
Experimental data and script files required for generating each figure can be found in the ZENODO repository, accessible via the following link: \href{https://doi.org/10.5281/zenodo.8072538}{https://doi.org/10.5281/zenodo.8072538}~\cite{zenodo_data}.


\section*{Acknowledgements}
 This work was supported by S\~{a}o Paulo Research Foundation (FAPESP) through grants 
19/09738-9, 
20/15786-3, 
19/01402-1, 
18/15577-5, 
18/15580-6, 
18/25339-4, 
22/07719-0, 
Coordena\c{c}\~{a}o de Aperfei\c{c}oamento de Pessoal de N\'{i}vel Superior - Brasil (CAPES) (Finance Code 001), the European Research Council (ERC CoG Q-ECHOS, 101001005), and by the Netherlands Organization for Scientific Research (NWO/OCW), as part of the Frontiers of Nanoscience program, as well as through Vrij Programma (680-92-18-04). We thank Prof. Fanny Béron and Prof. Kleber Pirota for providing access to the VNA used for OMIT/OMIA experiments and Dr. Felipe Santos for helpful comments.

\section*{Author contributions}
A.G.P, P.V.P. and T.P.M.A. devised and planned the experiment; R.B. fabricated the samples with design help from S.G. and T.P.M.A; A.G.P, P.V.P. performed measurements, data analysis and FEM simulations with contributions from G.S.W and T.P.M.A; S.G., G.S.W. and T.P.M.A. supervised the project. All authors contributed to the discussions and preparation of the manuscript.

\section*{Competing interests}
The authors declare no competing interests.

\section*{References}

\begin{thebibliography}{50}%
\makeatletter
\providecommand \@ifxundefined [1]{%
 \@ifx{#1\undefined}
}%
\providecommand \@ifnum [1]{%
 \ifnum #1\expandafter \@firstoftwo
 \else \expandafter \@secondoftwo
 \fi
}%
\providecommand \@ifx [1]{%
 \ifx #1\expandafter \@firstoftwo
 \else \expandafter \@secondoftwo
 \fi
}%
\providecommand \natexlab [1]{#1}%
\providecommand \enquote  [1]{``#1''}%
\providecommand \bibnamefont  [1]{#1}%
\providecommand \bibfnamefont [1]{#1}%
\providecommand \citenamefont [1]{#1}%
\providecommand \href@noop [0]{\@secondoftwo}%
\providecommand \href [0]{\begingroup \@sanitize@url \@href}%
\providecommand \@href[1]{\@@startlink{#1}\@@href}%
\providecommand \@@href[1]{\endgroup#1\@@endlink}%
\providecommand \@sanitize@url [0]{\catcode `\\12\catcode `\$12\catcode
  `\&12\catcode `\#12\catcode `\^12\catcode `\_12\catcode `\%12\relax}%
\providecommand \@@startlink[1]{}%
\providecommand \@@endlink[0]{}%
\providecommand \url  [0]{\begingroup\@sanitize@url \@url }%
\providecommand \@url [1]{\endgroup\@href {#1}{\urlprefix }}%
\providecommand \urlprefix  [0]{URL }%
\providecommand \Eprint [0]{\href }%
\providecommand \doibase [0]{http://dx.doi.org/}%
\providecommand \selectlanguage [0]{\@gobble}%
\providecommand \bibinfo  [0]{\@secondoftwo}%
\providecommand \bibfield  [0]{\@secondoftwo}%
\providecommand \translation [1]{[#1]}%
\providecommand \BibitemOpen [0]{}%
\providecommand \bibitemStop [0]{}%
\providecommand \bibitemNoStop [0]{.\EOS\space}%
\providecommand \EOS [0]{\spacefactor3000\relax}%
\providecommand \BibitemShut  [1]{\csname bibitem#1\endcsname}%
\let\auto@bib@innerbib\@empty
\bibitem [{\citenamefont {Wu}\ \emph {et~al.}(2014)\citenamefont {Wu},
  \citenamefont {Hryciw}, \citenamefont {Healey}, \citenamefont {Lake},
  \citenamefont {Jayakumar}, \citenamefont {Freeman}, \citenamefont {Davis},\
  and\ \citenamefont {Barclay}}]{wu_dissipative_2014}%
  \BibitemOpen
  \bibfield  {author} {\bibinfo {author} {\bibfnamefont {M.}~\bibnamefont
  {Wu}}, \bibinfo {author} {\bibfnamefont {A.~C.}\ \bibnamefont {Hryciw}},
  \bibinfo {author} {\bibfnamefont {C.}~\bibnamefont {Healey}}, \bibinfo
  {author} {\bibfnamefont {D.~P.}\ \bibnamefont {Lake}}, \bibinfo {author}
  {\bibfnamefont {H.}~\bibnamefont {Jayakumar}}, \bibinfo {author}
  {\bibfnamefont {M.~R.}\ \bibnamefont {Freeman}}, \bibinfo {author}
  {\bibfnamefont {J.~P.}\ \bibnamefont {Davis}}, \ and\ \bibinfo {author}
  {\bibfnamefont {P.~E.}\ \bibnamefont {Barclay}},\ }\href {\doibase
  10.1103/PhysRevX.4.021052} {\bibfield  {journal} {\bibinfo  {journal}
  {Physical Review X}\ }\textbf {\bibinfo {volume} {4}},\ \bibinfo {pages}
  {021052} (\bibinfo {year} {2014})}\BibitemShut {NoStop}%
\bibitem [{\citenamefont {Liu}\ \emph {et~al.}(2020)\citenamefont {Liu},
  \citenamefont {Pagliano}, \citenamefont {van Veldhoven}, \citenamefont
  {Pogoretskiy}, \citenamefont {Jiao},\ and\ \citenamefont
  {Fiore}}]{liu_integrated_2020}%
  \BibitemOpen
  \bibfield  {author} {\bibinfo {author} {\bibfnamefont {T.}~\bibnamefont
  {Liu}}, \bibinfo {author} {\bibfnamefont {F.}~\bibnamefont {Pagliano}},
  \bibinfo {author} {\bibfnamefont {R.}~\bibnamefont {van Veldhoven}}, \bibinfo
  {author} {\bibfnamefont {V.}~\bibnamefont {Pogoretskiy}}, \bibinfo {author}
  {\bibfnamefont {Y.}~\bibnamefont {Jiao}}, \ and\ \bibinfo {author}
  {\bibfnamefont {A.}~\bibnamefont {Fiore}},\ }\href {\doibase
  10.1038/s41467-020-16269-7} {\bibfield  {journal} {\bibinfo  {journal}
  {Nature Communications}\ }\textbf {\bibinfo {volume} {11}},\ \bibinfo {pages}
  {2407} (\bibinfo {year} {2020})}\BibitemShut {NoStop}%
\bibitem [{\citenamefont {Huber}\ \emph {et~al.}(2020)\citenamefont {Huber},
  \citenamefont {Rastelli}, \citenamefont {Seitner}, \citenamefont {Kölbl},
  \citenamefont {Belzig}, \citenamefont {Dykman},\ and\ \citenamefont
  {Weig}}]{huber_spectral_2020}%
  \BibitemOpen
  \bibfield  {author} {\bibinfo {author} {\bibfnamefont {J.~S.}\ \bibnamefont
  {Huber}}, \bibinfo {author} {\bibfnamefont {G.}~\bibnamefont {Rastelli}},
  \bibinfo {author} {\bibfnamefont {M.~J.}\ \bibnamefont {Seitner}}, \bibinfo
  {author} {\bibfnamefont {J.}~\bibnamefont {Kölbl}}, \bibinfo {author}
  {\bibfnamefont {W.}~\bibnamefont {Belzig}}, \bibinfo {author} {\bibfnamefont
  {M.~I.}\ \bibnamefont {Dykman}}, \ and\ \bibinfo {author} {\bibfnamefont
  {E.~M.}\ \bibnamefont {Weig}},\ }\href {\doibase 10.1103/PhysRevX.10.021066}
  {\bibfield  {journal} {\bibinfo  {journal} {Physical Review X}\ }\textbf
  {\bibinfo {volume} {10}},\ \bibinfo {pages} {021066} (\bibinfo {year}
  {2020})}\BibitemShut {NoStop}%
\bibitem [{\citenamefont {Rodrigues}\ \emph {et~al.}(2021)\citenamefont
  {Rodrigues}, \citenamefont {Kersul}, \citenamefont {Primo}, \citenamefont
  {Lipson}, \citenamefont {Alegre},\ and\ \citenamefont
  {Wiederhecker}}]{rodrigues_optomechanical_2021}%
  \BibitemOpen
  \bibfield  {author} {\bibinfo {author} {\bibfnamefont {C.~C.}\ \bibnamefont
  {Rodrigues}}, \bibinfo {author} {\bibfnamefont {C.~M.}\ \bibnamefont
  {Kersul}}, \bibinfo {author} {\bibfnamefont {A.~G.}\ \bibnamefont {Primo}},
  \bibinfo {author} {\bibfnamefont {M.}~\bibnamefont {Lipson}}, \bibinfo
  {author} {\bibfnamefont {T.~P.~M.}\ \bibnamefont {Alegre}}, \ and\ \bibinfo
  {author} {\bibfnamefont {G.~S.}\ \bibnamefont {Wiederhecker}},\ }\href
  {\doibase 10.1038/s41467-021-25884-x} {\bibfield  {journal} {\bibinfo
  {journal} {Nature Communications}\ }\textbf {\bibinfo {volume} {12}},\
  \bibinfo {pages} {5625} (\bibinfo {year} {2021})}\BibitemShut {NoStop}%
\bibitem [{\citenamefont {Zhang}\ \emph {et~al.}(2012)\citenamefont {Zhang},
  \citenamefont {Wiederhecker}, \citenamefont {Manipatruni}, \citenamefont
  {Barnard}, \citenamefont {McEuen},\ and\ \citenamefont
  {Lipson}}]{zhang_synchronization_2012}%
  \BibitemOpen
  \bibfield  {author} {\bibinfo {author} {\bibfnamefont {M.}~\bibnamefont
  {Zhang}}, \bibinfo {author} {\bibfnamefont {G.~S.}\ \bibnamefont
  {Wiederhecker}}, \bibinfo {author} {\bibfnamefont {S.}~\bibnamefont
  {Manipatruni}}, \bibinfo {author} {\bibfnamefont {A.}~\bibnamefont
  {Barnard}}, \bibinfo {author} {\bibfnamefont {P.}~\bibnamefont {McEuen}}, \
  and\ \bibinfo {author} {\bibfnamefont {M.}~\bibnamefont {Lipson}},\ }\href
  {\doibase 10.1103/PhysRevLett.109.233906} {\bibfield  {journal} {\bibinfo
  {journal} {Physical Review Letters}\ }\textbf {\bibinfo {volume} {109}},\
  \bibinfo {pages} {233906} (\bibinfo {year} {2012})}\BibitemShut {NoStop}%
\bibitem [{\citenamefont {Aspelmeyer}\ \emph {et~al.}(2014)\citenamefont
  {Aspelmeyer}, \citenamefont {Kippenberg},\ and\ \citenamefont
  {Marquardt}}]{aspelmeyer_cavity_2014}%
  \BibitemOpen
  \bibfield  {author} {\bibinfo {author} {\bibfnamefont {M.}~\bibnamefont
  {Aspelmeyer}}, \bibinfo {author} {\bibfnamefont {T.~J.}\ \bibnamefont
  {Kippenberg}}, \ and\ \bibinfo {author} {\bibfnamefont {F.}~\bibnamefont
  {Marquardt}},\ }\href {\doibase 10.1103/RevModPhys.86.1391} {\bibfield
  {journal} {\bibinfo  {journal} {Reviews of Modern Physics}\ }\textbf
  {\bibinfo {volume} {86}},\ \bibinfo {pages} {1391} (\bibinfo {year}
  {2014})}\BibitemShut {NoStop}%
\bibitem [{\citenamefont {Wiederhecker}\ \emph {et~al.}(2019)\citenamefont
  {Wiederhecker}, \citenamefont {Dainese},\ and\ \citenamefont
  {Mayer~Alegre}}]{wiederhecker_brillouin_2019}%
  \BibitemOpen
  \bibfield  {author} {\bibinfo {author} {\bibfnamefont {G.~S.}\ \bibnamefont
  {Wiederhecker}}, \bibinfo {author} {\bibfnamefont {P.}~\bibnamefont
  {Dainese}}, \ and\ \bibinfo {author} {\bibfnamefont {T.~P.}\ \bibnamefont
  {Mayer~Alegre}},\ }\href {\doibase 10.1063/1.5088169} {\bibfield  {journal}
  {\bibinfo  {journal} {APL Photonics}\ }\textbf {\bibinfo {volume} {4}},\
  \bibinfo {pages} {071101} (\bibinfo {year} {2019})}\BibitemShut {NoStop}%
\bibitem [{\citenamefont {Kippenberg}\ and\ \citenamefont
  {Vahala}(2008)}]{kippenberg_cavity_2008}%
  \BibitemOpen
  \bibfield  {author} {\bibinfo {author} {\bibfnamefont {T.~J.}\ \bibnamefont
  {Kippenberg}}\ and\ \bibinfo {author} {\bibfnamefont {K.~J.}\ \bibnamefont
  {Vahala}},\ }\href {\doibase 10.1126/science.1156032} {\bibfield  {journal}
  {\bibinfo  {journal} {Science}\ }\textbf {\bibinfo {volume} {321}},\ \bibinfo
  {pages} {1172} (\bibinfo {year} {2008})}\BibitemShut {NoStop}%
\bibitem [{\citenamefont {He}\ \emph {et~al.}(2020)\citenamefont {He},
  \citenamefont {Harris}, \citenamefont {Baker}, \citenamefont {Sawadsky},
  \citenamefont {Sfendla}, \citenamefont {Sachkou}, \citenamefont {Forstner},\
  and\ \citenamefont {Bowen}}]{he_strong_2020}%
  \BibitemOpen
  \bibfield  {author} {\bibinfo {author} {\bibfnamefont {X.}~\bibnamefont
  {He}}, \bibinfo {author} {\bibfnamefont {G.~I.}\ \bibnamefont {Harris}},
  \bibinfo {author} {\bibfnamefont {C.~G.}\ \bibnamefont {Baker}}, \bibinfo
  {author} {\bibfnamefont {A.}~\bibnamefont {Sawadsky}}, \bibinfo {author}
  {\bibfnamefont {Y.~L.}\ \bibnamefont {Sfendla}}, \bibinfo {author}
  {\bibfnamefont {Y.~P.}\ \bibnamefont {Sachkou}}, \bibinfo {author}
  {\bibfnamefont {S.}~\bibnamefont {Forstner}}, \ and\ \bibinfo {author}
  {\bibfnamefont {W.~P.}\ \bibnamefont {Bowen}},\ }\href {\doibase
  10.1038/s41567-020-0785-0} {\bibfield  {journal} {\bibinfo  {journal} {Nature
  Physics}\ }\textbf {\bibinfo {volume} {16}},\ \bibinfo {pages} {417}
  (\bibinfo {year} {2020})}\BibitemShut {NoStop}%
\bibitem [{\citenamefont {Jiang}\ \emph {et~al.}(2022)\citenamefont {Jiang},
  \citenamefont {Mayor}, \citenamefont {Malik}, \citenamefont {Van~Laer},
  \citenamefont {McKenna}, \citenamefont {Patel}, \citenamefont {Witmer},\ and\
  \citenamefont {Safavi-Naeini}}]{jiang_optically_2022}%
  \BibitemOpen
  \bibfield  {author} {\bibinfo {author} {\bibfnamefont {W.}~\bibnamefont
  {Jiang}}, \bibinfo {author} {\bibfnamefont {F.~M.}\ \bibnamefont {Mayor}},
  \bibinfo {author} {\bibfnamefont {S.}~\bibnamefont {Malik}}, \bibinfo
  {author} {\bibfnamefont {R.}~\bibnamefont {Van~Laer}}, \bibinfo {author}
  {\bibfnamefont {T.~P.}\ \bibnamefont {McKenna}}, \bibinfo {author}
  {\bibfnamefont {R.~N.}\ \bibnamefont {Patel}}, \bibinfo {author}
  {\bibfnamefont {J.~D.}\ \bibnamefont {Witmer}}, \ and\ \bibinfo {author}
  {\bibfnamefont {A.~H.}\ \bibnamefont {Safavi-Naeini}},\ }\href {\doibase
  10.48550/arXiv.2210.10739} {\enquote {\bibinfo {title} {Optically heralded
  microwave photons},}\ } (\bibinfo {year} {2022})\BibitemShut {NoStop}%
\bibitem [{\citenamefont {Mirhosseini}\ \emph {et~al.}(2020)\citenamefont
  {Mirhosseini}, \citenamefont {Sipahigil}, \citenamefont {Kalaee},\ and\
  \citenamefont {Painter}}]{mirhosseini_superconducting_2020}%
  \BibitemOpen
  \bibfield  {author} {\bibinfo {author} {\bibfnamefont {M.}~\bibnamefont
  {Mirhosseini}}, \bibinfo {author} {\bibfnamefont {A.}~\bibnamefont
  {Sipahigil}}, \bibinfo {author} {\bibfnamefont {M.}~\bibnamefont {Kalaee}}, \
  and\ \bibinfo {author} {\bibfnamefont {O.}~\bibnamefont {Painter}},\ }\href
  {\doibase 10.1038/s41586-020-3038-6} {\bibfield  {journal} {\bibinfo
  {journal} {Nature}\ }\textbf {\bibinfo {volume} {588}},\ \bibinfo {pages}
  {599} (\bibinfo {year} {2020})}\BibitemShut {NoStop}%
\bibitem [{\citenamefont {Forsch}\ \emph {et~al.}(2020)\citenamefont {Forsch},
  \citenamefont {Stockill}, \citenamefont {Wallucks}, \citenamefont
  {Marinković}, \citenamefont {Gärtner}, \citenamefont {Norte}, \citenamefont
  {van Otten}, \citenamefont {Fiore}, \citenamefont {Srinivasan},\ and\
  \citenamefont {Gröblacher}}]{forsch_microwave--optics_2020}%
  \BibitemOpen
  \bibfield  {author} {\bibinfo {author} {\bibfnamefont {M.}~\bibnamefont
  {Forsch}}, \bibinfo {author} {\bibfnamefont {R.}~\bibnamefont {Stockill}},
  \bibinfo {author} {\bibfnamefont {A.}~\bibnamefont {Wallucks}}, \bibinfo
  {author} {\bibfnamefont {I.}~\bibnamefont {Marinković}}, \bibinfo {author}
  {\bibfnamefont {C.}~\bibnamefont {Gärtner}}, \bibinfo {author}
  {\bibfnamefont {R.~A.}\ \bibnamefont {Norte}}, \bibinfo {author}
  {\bibfnamefont {F.}~\bibnamefont {van Otten}}, \bibinfo {author}
  {\bibfnamefont {A.}~\bibnamefont {Fiore}}, \bibinfo {author} {\bibfnamefont
  {K.}~\bibnamefont {Srinivasan}}, \ and\ \bibinfo {author} {\bibfnamefont
  {S.}~\bibnamefont {Gröblacher}},\ }\href {\doibase
  10.1038/s41567-019-0673-7} {\bibfield  {journal} {\bibinfo  {journal} {Nature
  Physics}\ }\textbf {\bibinfo {volume} {16}},\ \bibinfo {pages} {69} (\bibinfo
  {year} {2020})}\BibitemShut {NoStop}%
\bibitem [{\citenamefont {Jiang}\ \emph {et~al.}(2020)\citenamefont {Jiang},
  \citenamefont {Sarabalis}, \citenamefont {Dahmani}, \citenamefont {Patel},
  \citenamefont {Mayor}, \citenamefont {McKenna}, \citenamefont {Van~Laer},\
  and\ \citenamefont {Safavi-Naeini}}]{jiang_efficient_2020}%
  \BibitemOpen
  \bibfield  {author} {\bibinfo {author} {\bibfnamefont {W.}~\bibnamefont
  {Jiang}}, \bibinfo {author} {\bibfnamefont {C.~J.}\ \bibnamefont
  {Sarabalis}}, \bibinfo {author} {\bibfnamefont {Y.~D.}\ \bibnamefont
  {Dahmani}}, \bibinfo {author} {\bibfnamefont {R.~N.}\ \bibnamefont {Patel}},
  \bibinfo {author} {\bibfnamefont {F.~M.}\ \bibnamefont {Mayor}}, \bibinfo
  {author} {\bibfnamefont {T.~P.}\ \bibnamefont {McKenna}}, \bibinfo {author}
  {\bibfnamefont {R.}~\bibnamefont {Van~Laer}}, \ and\ \bibinfo {author}
  {\bibfnamefont {A.~H.}\ \bibnamefont {Safavi-Naeini}},\ }\href {\doibase
  10.1038/s41467-020-14863-3} {\bibfield  {journal} {\bibinfo  {journal}
  {Nature Communications}\ }\textbf {\bibinfo {volume} {11}},\ \bibinfo {pages}
  {1166} (\bibinfo {year} {2020})}\BibitemShut {NoStop}%
\bibitem [{\citenamefont {Rueda}\ \emph {et~al.}(2016)\citenamefont {Rueda},
  \citenamefont {Sedlmeir}, \citenamefont {Collodo}, \citenamefont {Vogl},
  \citenamefont {Stiller}, \citenamefont {Schunk}, \citenamefont {Strekalov},
  \citenamefont {Marquardt}, \citenamefont {Fink}, \citenamefont {Painter},
  \citenamefont {Leuchs},\ and\ \citenamefont
  {Schwefel}}]{rueda_efficient_2016}%
  \BibitemOpen
  \bibfield  {author} {\bibinfo {author} {\bibfnamefont {A.}~\bibnamefont
  {Rueda}}, \bibinfo {author} {\bibfnamefont {F.}~\bibnamefont {Sedlmeir}},
  \bibinfo {author} {\bibfnamefont {M.~C.}\ \bibnamefont {Collodo}}, \bibinfo
  {author} {\bibfnamefont {U.}~\bibnamefont {Vogl}}, \bibinfo {author}
  {\bibfnamefont {B.}~\bibnamefont {Stiller}}, \bibinfo {author} {\bibfnamefont
  {G.}~\bibnamefont {Schunk}}, \bibinfo {author} {\bibfnamefont {D.~V.}\
  \bibnamefont {Strekalov}}, \bibinfo {author} {\bibfnamefont {C.}~\bibnamefont
  {Marquardt}}, \bibinfo {author} {\bibfnamefont {J.~M.}\ \bibnamefont {Fink}},
  \bibinfo {author} {\bibfnamefont {O.}~\bibnamefont {Painter}}, \bibinfo
  {author} {\bibfnamefont {G.}~\bibnamefont {Leuchs}}, \ and\ \bibinfo {author}
  {\bibfnamefont {H.~G.~L.}\ \bibnamefont {Schwefel}},\ }\href {\doibase
  10.1364/OPTICA.3.000597} {\bibfield  {journal} {\bibinfo  {journal} {Optica}\
  }\textbf {\bibinfo {volume} {3}},\ \bibinfo {pages} {597} (\bibinfo {year}
  {2016})}\BibitemShut {NoStop}%
\bibitem [{\citenamefont {Barzanjeh}\ \emph {et~al.}(2022)\citenamefont
  {Barzanjeh}, \citenamefont {Xuereb}, \citenamefont {Gröblacher},
  \citenamefont {Paternostro}, \citenamefont {Regal},\ and\ \citenamefont
  {Weig}}]{barzanjeh_optomechanics_2022}%
  \BibitemOpen
  \bibfield  {author} {\bibinfo {author} {\bibfnamefont {S.}~\bibnamefont
  {Barzanjeh}}, \bibinfo {author} {\bibfnamefont {A.}~\bibnamefont {Xuereb}},
  \bibinfo {author} {\bibfnamefont {S.}~\bibnamefont {Gröblacher}}, \bibinfo
  {author} {\bibfnamefont {M.}~\bibnamefont {Paternostro}}, \bibinfo {author}
  {\bibfnamefont {C.~A.}\ \bibnamefont {Regal}}, \ and\ \bibinfo {author}
  {\bibfnamefont {E.~M.}\ \bibnamefont {Weig}},\ }\href {\doibase
  10.1038/s41567-021-01402-0} {\bibfield  {journal} {\bibinfo  {journal}
  {Nature Physics}\ }\textbf {\bibinfo {volume} {18}},\ \bibinfo {pages} {15}
  (\bibinfo {year} {2022})}\BibitemShut {NoStop}%
\bibitem [{\citenamefont {McKenna}\ \emph {et~al.}(2020)\citenamefont
  {McKenna}, \citenamefont {Witmer}, \citenamefont {Patel}, \citenamefont
  {Jiang}, \citenamefont {Van~Laer}, \citenamefont {Arrangoiz-Arriola},
  \citenamefont {Wollack}, \citenamefont {Herrmann},\ and\ \citenamefont
  {Safavi-Naeini}}]{mckenna_cryogenic_2020}%
  \BibitemOpen
  \bibfield  {author} {\bibinfo {author} {\bibfnamefont {T.~P.}\ \bibnamefont
  {McKenna}}, \bibinfo {author} {\bibfnamefont {J.~D.}\ \bibnamefont {Witmer}},
  \bibinfo {author} {\bibfnamefont {R.~N.}\ \bibnamefont {Patel}}, \bibinfo
  {author} {\bibfnamefont {W.}~\bibnamefont {Jiang}}, \bibinfo {author}
  {\bibfnamefont {R.}~\bibnamefont {Van~Laer}}, \bibinfo {author}
  {\bibfnamefont {P.}~\bibnamefont {Arrangoiz-Arriola}}, \bibinfo {author}
  {\bibfnamefont {E.~A.}\ \bibnamefont {Wollack}}, \bibinfo {author}
  {\bibfnamefont {J.~F.}\ \bibnamefont {Herrmann}}, \ and\ \bibinfo {author}
  {\bibfnamefont {A.~H.}\ \bibnamefont {Safavi-Naeini}},\ }\href {\doibase
  10.1364/OPTICA.397235} {\bibfield  {journal} {\bibinfo  {journal} {Optica}\
  }\textbf {\bibinfo {volume} {7}},\ \bibinfo {pages} {1737} (\bibinfo {year}
  {2020})}\BibitemShut {NoStop}%
\bibitem [{\citenamefont {Han}\ \emph {et~al.}(2021)\citenamefont {Han},
  \citenamefont {Fu}, \citenamefont {Zou}, \citenamefont {Jiang},\ and\
  \citenamefont {Tang}}]{han_microwave-optical_2021}%
  \BibitemOpen
  \bibfield  {author} {\bibinfo {author} {\bibfnamefont {X.}~\bibnamefont
  {Han}}, \bibinfo {author} {\bibfnamefont {W.}~\bibnamefont {Fu}}, \bibinfo
  {author} {\bibfnamefont {C.-L.}\ \bibnamefont {Zou}}, \bibinfo {author}
  {\bibfnamefont {L.}~\bibnamefont {Jiang}}, \ and\ \bibinfo {author}
  {\bibfnamefont {H.~X.}\ \bibnamefont {Tang}},\ }\href {\doibase
  10.1364/OPTICA.425414} {\bibfield  {journal} {\bibinfo  {journal} {Optica}\
  }\textbf {\bibinfo {volume} {8}},\ \bibinfo {pages} {1050} (\bibinfo {year}
  {2021})}\BibitemShut {NoStop}%
\bibitem [{\citenamefont {Shao}\ \emph {et~al.}(2019)\citenamefont {Shao},
  \citenamefont {Yu}, \citenamefont {Maity}, \citenamefont {Sinclair},
  \citenamefont {Zheng}, \citenamefont {Chia}, \citenamefont {Shams-Ansari},
  \citenamefont {Wang}, \citenamefont {Zhang}, \citenamefont {Lai},\ and\
  \citenamefont {Lončar}}]{shao_microwave--optical_2019}%
  \BibitemOpen
  \bibfield  {author} {\bibinfo {author} {\bibfnamefont {L.}~\bibnamefont
  {Shao}}, \bibinfo {author} {\bibfnamefont {M.}~\bibnamefont {Yu}}, \bibinfo
  {author} {\bibfnamefont {S.}~\bibnamefont {Maity}}, \bibinfo {author}
  {\bibfnamefont {N.}~\bibnamefont {Sinclair}}, \bibinfo {author}
  {\bibfnamefont {L.}~\bibnamefont {Zheng}}, \bibinfo {author} {\bibfnamefont
  {C.}~\bibnamefont {Chia}}, \bibinfo {author} {\bibfnamefont {A.}~\bibnamefont
  {Shams-Ansari}}, \bibinfo {author} {\bibfnamefont {C.}~\bibnamefont {Wang}},
  \bibinfo {author} {\bibfnamefont {M.}~\bibnamefont {Zhang}}, \bibinfo
  {author} {\bibfnamefont {K.}~\bibnamefont {Lai}}, \ and\ \bibinfo {author}
  {\bibfnamefont {M.}~\bibnamefont {Lončar}},\ }\href {\doibase
  10.1364/OPTICA.6.001498} {\bibfield  {journal} {\bibinfo  {journal} {Optica}\
  }\textbf {\bibinfo {volume} {6}},\ \bibinfo {pages} {1498} (\bibinfo {year}
  {2019})}\BibitemShut {NoStop}%
\bibitem [{\citenamefont {Hönl}\ \emph {et~al.}(2022)\citenamefont {Hönl},
  \citenamefont {Popoff}, \citenamefont {Caimi}, \citenamefont {Beccari},
  \citenamefont {Kippenberg},\ and\ \citenamefont
  {Seidler}}]{honl_microwave--optical_2022}%
  \BibitemOpen
  \bibfield  {author} {\bibinfo {author} {\bibfnamefont {S.}~\bibnamefont
  {Hönl}}, \bibinfo {author} {\bibfnamefont {Y.}~\bibnamefont {Popoff}},
  \bibinfo {author} {\bibfnamefont {D.}~\bibnamefont {Caimi}}, \bibinfo
  {author} {\bibfnamefont {A.}~\bibnamefont {Beccari}}, \bibinfo {author}
  {\bibfnamefont {T.~J.}\ \bibnamefont {Kippenberg}}, \ and\ \bibinfo {author}
  {\bibfnamefont {P.}~\bibnamefont {Seidler}},\ }\href {\doibase
  10.1038/s41467-022-28670-5} {\bibfield  {journal} {\bibinfo  {journal}
  {Nature Communications}\ }\textbf {\bibinfo {volume} {13}},\ \bibinfo {pages}
  {2065} (\bibinfo {year} {2022})}\BibitemShut {NoStop}%
\bibitem [{\citenamefont {Wallucks}\ \emph {et~al.}(2020)\citenamefont
  {Wallucks}, \citenamefont {Marinković}, \citenamefont {Hensen},
  \citenamefont {Stockill},\ and\ \citenamefont
  {Gröblacher}}]{wallucks_quantum_2020}%
  \BibitemOpen
  \bibfield  {author} {\bibinfo {author} {\bibfnamefont {A.}~\bibnamefont
  {Wallucks}}, \bibinfo {author} {\bibfnamefont {I.}~\bibnamefont
  {Marinković}}, \bibinfo {author} {\bibfnamefont {B.}~\bibnamefont {Hensen}},
  \bibinfo {author} {\bibfnamefont {R.}~\bibnamefont {Stockill}}, \ and\
  \bibinfo {author} {\bibfnamefont {S.}~\bibnamefont {Gröblacher}},\ }\href
  {\doibase 10.1038/s41567-020-0891-z} {\bibfield  {journal} {\bibinfo
  {journal} {Nature Physics}\ }\textbf {\bibinfo {volume} {16}},\ \bibinfo
  {pages} {772} (\bibinfo {year} {2020})}\BibitemShut {NoStop}%
\bibitem [{\citenamefont {Stiller}\ \emph {et~al.}(2020)\citenamefont
  {Stiller}, \citenamefont {Merklein}, \citenamefont {Wolff}, \citenamefont
  {Vu}, \citenamefont {Ma}, \citenamefont {Madden},\ and\ \citenamefont
  {Eggleton}}]{stiller_coherently_2020}%
  \BibitemOpen
  \bibfield  {author} {\bibinfo {author} {\bibfnamefont {B.}~\bibnamefont
  {Stiller}}, \bibinfo {author} {\bibfnamefont {M.}~\bibnamefont {Merklein}},
  \bibinfo {author} {\bibfnamefont {C.}~\bibnamefont {Wolff}}, \bibinfo
  {author} {\bibfnamefont {K.}~\bibnamefont {Vu}}, \bibinfo {author}
  {\bibfnamefont {P.}~\bibnamefont {Ma}}, \bibinfo {author} {\bibfnamefont
  {S.~J.}\ \bibnamefont {Madden}}, \ and\ \bibinfo {author} {\bibfnamefont
  {B.~J.}\ \bibnamefont {Eggleton}},\ }\href {\doibase 10.1364/OPTICA.386535}
  {\bibfield  {journal} {\bibinfo  {journal} {Optica}\ }\textbf {\bibinfo
  {volume} {7}},\ \bibinfo {pages} {492} (\bibinfo {year} {2020})}\BibitemShut
  {NoStop}%
\bibitem [{\citenamefont {Shomroni}\ \emph {et~al.}(2019)\citenamefont
  {Shomroni}, \citenamefont {Qiu}, \citenamefont {Malz}, \citenamefont
  {Nunnenkamp},\ and\ \citenamefont {Kippenberg}}]{shomroni_optical_2019}%
  \BibitemOpen
  \bibfield  {author} {\bibinfo {author} {\bibfnamefont {I.}~\bibnamefont
  {Shomroni}}, \bibinfo {author} {\bibfnamefont {L.}~\bibnamefont {Qiu}},
  \bibinfo {author} {\bibfnamefont {D.}~\bibnamefont {Malz}}, \bibinfo {author}
  {\bibfnamefont {A.}~\bibnamefont {Nunnenkamp}}, \ and\ \bibinfo {author}
  {\bibfnamefont {T.~J.}\ \bibnamefont {Kippenberg}},\ }\href {\doibase
  10.1038/s41467-019-10024-3} {\bibfield  {journal} {\bibinfo  {journal}
  {Nature Communications}\ }\textbf {\bibinfo {volume} {10}},\ \bibinfo {pages}
  {2086} (\bibinfo {year} {2019})}\BibitemShut {NoStop}%
\bibitem [{\citenamefont {Chan}\ \emph {et~al.}(2011)\citenamefont {Chan},
  \citenamefont {Alegre}, \citenamefont {Safavi-Naeini}, \citenamefont {Hill},
  \citenamefont {Krause}, \citenamefont {Gröblacher}, \citenamefont
  {Aspelmeyer},\ and\ \citenamefont {Painter}}]{chan_laser_2011}%
  \BibitemOpen
  \bibfield  {author} {\bibinfo {author} {\bibfnamefont {J.}~\bibnamefont
  {Chan}}, \bibinfo {author} {\bibfnamefont {T.~P.~M.}\ \bibnamefont {Alegre}},
  \bibinfo {author} {\bibfnamefont {A.~H.}\ \bibnamefont {Safavi-Naeini}},
  \bibinfo {author} {\bibfnamefont {J.~T.}\ \bibnamefont {Hill}}, \bibinfo
  {author} {\bibfnamefont {A.}~\bibnamefont {Krause}}, \bibinfo {author}
  {\bibfnamefont {S.}~\bibnamefont {Gröblacher}}, \bibinfo {author}
  {\bibfnamefont {M.}~\bibnamefont {Aspelmeyer}}, \ and\ \bibinfo {author}
  {\bibfnamefont {O.}~\bibnamefont {Painter}},\ }\href {\doibase
  10.1038/nature10461} {\bibfield  {journal} {\bibinfo  {journal} {Nature}\
  }\textbf {\bibinfo {volume} {478}},\ \bibinfo {pages} {89} (\bibinfo {year}
  {2011})}\BibitemShut {NoStop}%
\bibitem [{\citenamefont {Fiaschi}\ \emph {et~al.}(2021)\citenamefont
  {Fiaschi}, \citenamefont {Hensen}, \citenamefont {Wallucks}, \citenamefont
  {Benevides}, \citenamefont {Li}, \citenamefont {Alegre},\ and\ \citenamefont
  {Gröblacher}}]{fiaschi_optomechanical_2021}%
  \BibitemOpen
  \bibfield  {author} {\bibinfo {author} {\bibfnamefont {N.}~\bibnamefont
  {Fiaschi}}, \bibinfo {author} {\bibfnamefont {B.}~\bibnamefont {Hensen}},
  \bibinfo {author} {\bibfnamefont {A.}~\bibnamefont {Wallucks}}, \bibinfo
  {author} {\bibfnamefont {R.}~\bibnamefont {Benevides}}, \bibinfo {author}
  {\bibfnamefont {J.}~\bibnamefont {Li}}, \bibinfo {author} {\bibfnamefont
  {T.~P.~M.}\ \bibnamefont {Alegre}}, \ and\ \bibinfo {author} {\bibfnamefont
  {S.}~\bibnamefont {Gröblacher}},\ }\href {\doibase
  10.1038/s41566-021-00866-z} {\bibfield  {journal} {\bibinfo  {journal}
  {Nature Photonics}\ }\textbf {\bibinfo {volume} {15}},\ \bibinfo {pages}
  {817} (\bibinfo {year} {2021})}\BibitemShut {NoStop}%
\bibitem [{\citenamefont {Elste}\ \emph {et~al.}(2009)\citenamefont {Elste},
  \citenamefont {Girvin},\ and\ \citenamefont {Clerk}}]{elste_quantum_2009}%
  \BibitemOpen
  \bibfield  {author} {\bibinfo {author} {\bibfnamefont {F.}~\bibnamefont
  {Elste}}, \bibinfo {author} {\bibfnamefont {S.~M.}\ \bibnamefont {Girvin}}, \
  and\ \bibinfo {author} {\bibfnamefont {A.~A.}\ \bibnamefont {Clerk}},\ }\href
  {\doibase 10.1103/PhysRevLett.102.207209} {\bibfield  {journal} {\bibinfo
  {journal} {Physical Review Letters}\ }\textbf {\bibinfo {volume} {102}},\
  \bibinfo {pages} {207209} (\bibinfo {year} {2009})}\BibitemShut {NoStop}%
\bibitem [{\citenamefont {Weiss}\ and\ \citenamefont
  {Nunnenkamp}(2013)}]{weiss_quantum_2013}%
  \BibitemOpen
  \bibfield  {author} {\bibinfo {author} {\bibfnamefont {T.}~\bibnamefont
  {Weiss}}\ and\ \bibinfo {author} {\bibfnamefont {A.}~\bibnamefont
  {Nunnenkamp}},\ }\href {\doibase 10.1103/PhysRevA.88.023850} {\bibfield
  {journal} {\bibinfo  {journal} {Physical Review A}\ }\textbf {\bibinfo
  {volume} {88}},\ \bibinfo {pages} {023850} (\bibinfo {year}
  {2013})}\BibitemShut {NoStop}%
\bibitem [{\citenamefont {Weiss}\ \emph {et~al.}(2013)\citenamefont {Weiss},
  \citenamefont {Bruder},\ and\ \citenamefont
  {Nunnenkamp}}]{weiss_strong-coupling_2013}%
  \BibitemOpen
  \bibfield  {author} {\bibinfo {author} {\bibfnamefont {T.}~\bibnamefont
  {Weiss}}, \bibinfo {author} {\bibfnamefont {C.}~\bibnamefont {Bruder}}, \
  and\ \bibinfo {author} {\bibfnamefont {A.}~\bibnamefont {Nunnenkamp}},\
  }\href {\doibase 10.1088/1367-2630/15/4/045017} {\bibfield  {journal}
  {\bibinfo  {journal} {New Journal of Physics}\ }\textbf {\bibinfo {volume}
  {15}},\ \bibinfo {pages} {045017} (\bibinfo {year} {2013})}\BibitemShut
  {NoStop}%
\bibitem [{\citenamefont {Primo}\ \emph {et~al.}(2020)\citenamefont {Primo},
  \citenamefont {Carvalho}, \citenamefont {Kersul}, \citenamefont {Frateschi},
  \citenamefont {Wiederhecker},\ and\ \citenamefont
  {Alegre}}]{primo_quasinormal-mode_2020}%
  \BibitemOpen
  \bibfield  {author} {\bibinfo {author} {\bibfnamefont {A.~G.}\ \bibnamefont
  {Primo}}, \bibinfo {author} {\bibfnamefont {N.~C.}\ \bibnamefont {Carvalho}},
  \bibinfo {author} {\bibfnamefont {C.~M.}\ \bibnamefont {Kersul}}, \bibinfo
  {author} {\bibfnamefont {N.~C.}\ \bibnamefont {Frateschi}}, \bibinfo {author}
  {\bibfnamefont {G.~S.}\ \bibnamefont {Wiederhecker}}, \ and\ \bibinfo
  {author} {\bibfnamefont {T.~P.~M.}\ \bibnamefont {Alegre}},\ }\href {\doibase
  10.1103/PhysRevLett.125.233601} {\bibfield  {journal} {\bibinfo  {journal}
  {Physical Review Letters}\ }\textbf {\bibinfo {volume} {125}},\ \bibinfo
  {pages} {233601} (\bibinfo {year} {2020})}\BibitemShut {NoStop}%
\bibitem [{\citenamefont {Yanay}\ \emph {et~al.}(2016)\citenamefont {Yanay},
  \citenamefont {Sankey},\ and\ \citenamefont {Clerk}}]{yanay_quantum_2016}%
  \BibitemOpen
  \bibfield  {author} {\bibinfo {author} {\bibfnamefont {Y.}~\bibnamefont
  {Yanay}}, \bibinfo {author} {\bibfnamefont {J.~C.}\ \bibnamefont {Sankey}}, \
  and\ \bibinfo {author} {\bibfnamefont {A.~A.}\ \bibnamefont {Clerk}},\ }\href
  {\doibase 10.1103/PhysRevA.93.063809} {\bibfield  {journal} {\bibinfo
  {journal} {Physical Review A}\ }\textbf {\bibinfo {volume} {93}},\ \bibinfo
  {pages} {063809} (\bibinfo {year} {2016})}\BibitemShut {NoStop}%
\bibitem [{\citenamefont {Genes}\ \emph {et~al.}(2009)\citenamefont {Genes},
  \citenamefont {Ritsch},\ and\ \citenamefont
  {Vitali}}]{genes_micromechanical_2009}%
  \BibitemOpen
  \bibfield  {author} {\bibinfo {author} {\bibfnamefont {C.}~\bibnamefont
  {Genes}}, \bibinfo {author} {\bibfnamefont {H.}~\bibnamefont {Ritsch}}, \
  and\ \bibinfo {author} {\bibfnamefont {D.}~\bibnamefont {Vitali}},\ }\href
  {\doibase 10.1103/PhysRevA.80.061803} {\bibfield  {journal} {\bibinfo
  {journal} {Physical Review A}\ }\textbf {\bibinfo {volume} {80}},\ \bibinfo
  {pages} {061803} (\bibinfo {year} {2009})}\BibitemShut {NoStop}%
\bibitem [{\citenamefont {Huang}\ \emph {et~al.}(2018)\citenamefont {Huang},
  \citenamefont {Li}, \citenamefont {Chin}, \citenamefont {Cai}, \citenamefont
  {Gu}, \citenamefont {Karim}, \citenamefont {Wu}, \citenamefont {Chen},
  \citenamefont {Yang}, \citenamefont {Hao}, \citenamefont {Qiu},\ and\
  \citenamefont {Liu}}]{huang_dissipative_2018}%
  \BibitemOpen
  \bibfield  {author} {\bibinfo {author} {\bibfnamefont {J.~G.}\ \bibnamefont
  {Huang}}, \bibinfo {author} {\bibfnamefont {Y.}~\bibnamefont {Li}}, \bibinfo
  {author} {\bibfnamefont {L.~K.}\ \bibnamefont {Chin}}, \bibinfo {author}
  {\bibfnamefont {H.}~\bibnamefont {Cai}}, \bibinfo {author} {\bibfnamefont
  {Y.~D.}\ \bibnamefont {Gu}}, \bibinfo {author} {\bibfnamefont {M.~F.}\
  \bibnamefont {Karim}}, \bibinfo {author} {\bibfnamefont {J.~H.}\ \bibnamefont
  {Wu}}, \bibinfo {author} {\bibfnamefont {T.~N.}\ \bibnamefont {Chen}},
  \bibinfo {author} {\bibfnamefont {Z.~C.}\ \bibnamefont {Yang}}, \bibinfo
  {author} {\bibfnamefont {Y.~L.}\ \bibnamefont {Hao}}, \bibinfo {author}
  {\bibfnamefont {C.~W.}\ \bibnamefont {Qiu}}, \ and\ \bibinfo {author}
  {\bibfnamefont {A.~Q.}\ \bibnamefont {Liu}},\ }\href {\doibase
  10.1063/1.5009402} {\bibfield  {journal} {\bibinfo  {journal} {Applied
  Physics Letters}\ }\textbf {\bibinfo {volume} {112}},\ \bibinfo {pages}
  {051104} (\bibinfo {year} {2018})}\BibitemShut {NoStop}%
\bibitem [{\citenamefont {Li}\ \emph {et~al.}(2009)\citenamefont {Li},
  \citenamefont {Pernice},\ and\ \citenamefont {Tang}}]{li_reactive_2009}%
  \BibitemOpen
  \bibfield  {author} {\bibinfo {author} {\bibfnamefont {M.}~\bibnamefont
  {Li}}, \bibinfo {author} {\bibfnamefont {W.~H.~P.}\ \bibnamefont {Pernice}},
  \ and\ \bibinfo {author} {\bibfnamefont {H.~X.}\ \bibnamefont {Tang}},\
  }\href {\doibase 10.1103/PhysRevLett.103.223901} {\bibfield  {journal}
  {\bibinfo  {journal} {Physical Review Letters}\ }\textbf {\bibinfo {volume}
  {103}},\ \bibinfo {pages} {223901} (\bibinfo {year} {2009})}\BibitemShut
  {NoStop}%
\bibitem [{\citenamefont {Barnard}\ \emph {et~al.}(2019)\citenamefont
  {Barnard}, \citenamefont {Zhang}, \citenamefont {Wiederhecker}, \citenamefont
  {Lipson},\ and\ \citenamefont {McEuen}}]{barnard_real-time_2019}%
  \BibitemOpen
  \bibfield  {author} {\bibinfo {author} {\bibfnamefont {A.~W.}\ \bibnamefont
  {Barnard}}, \bibinfo {author} {\bibfnamefont {M.}~\bibnamefont {Zhang}},
  \bibinfo {author} {\bibfnamefont {G.~S.}\ \bibnamefont {Wiederhecker}},
  \bibinfo {author} {\bibfnamefont {M.}~\bibnamefont {Lipson}}, \ and\ \bibinfo
  {author} {\bibfnamefont {P.~L.}\ \bibnamefont {McEuen}},\ }\href {\doibase
  10.1038/s41586-018-0861-0} {\bibfield  {journal} {\bibinfo  {journal}
  {Nature}\ }\textbf {\bibinfo {volume} {566}},\ \bibinfo {pages} {89}
  (\bibinfo {year} {2019})}\BibitemShut {NoStop}%
\bibitem [{\citenamefont {Weis}\ \emph {et~al.}(2010)\citenamefont {Weis},
  \citenamefont {Rivière}, \citenamefont {Deléglise}, \citenamefont
  {Gavartin}, \citenamefont {Arcizet}, \citenamefont {Schliesser},\ and\
  \citenamefont {Kippenberg}}]{weis_optomechanically_2010}%
  \BibitemOpen
  \bibfield  {author} {\bibinfo {author} {\bibfnamefont {S.}~\bibnamefont
  {Weis}}, \bibinfo {author} {\bibfnamefont {R.}~\bibnamefont {Rivière}},
  \bibinfo {author} {\bibfnamefont {S.}~\bibnamefont {Deléglise}}, \bibinfo
  {author} {\bibfnamefont {E.}~\bibnamefont {Gavartin}}, \bibinfo {author}
  {\bibfnamefont {O.}~\bibnamefont {Arcizet}}, \bibinfo {author} {\bibfnamefont
  {A.}~\bibnamefont {Schliesser}}, \ and\ \bibinfo {author} {\bibfnamefont
  {T.~J.}\ \bibnamefont {Kippenberg}},\ }\href {\doibase
  10.1126/science.1195596} {\bibfield  {journal} {\bibinfo  {journal}
  {Science}\ }\textbf {\bibinfo {volume} {330}},\ \bibinfo {pages} {1520}
  (\bibinfo {year} {2010})}\BibitemShut {NoStop}%
\bibitem [{\citenamefont {Safavi-Naeini}\ \emph {et~al.}(2011)\citenamefont
  {Safavi-Naeini}, \citenamefont {Alegre}, \citenamefont {Chan}, \citenamefont
  {Eichenfield}, \citenamefont {Winger}, \citenamefont {Lin}, \citenamefont
  {Hill}, \citenamefont {Chang},\ and\ \citenamefont
  {Painter}}]{safavi-naeini_electromagnetically_2011}%
  \BibitemOpen
  \bibfield  {author} {\bibinfo {author} {\bibfnamefont {A.~H.}\ \bibnamefont
  {Safavi-Naeini}}, \bibinfo {author} {\bibfnamefont {T.~P.~M.}\ \bibnamefont
  {Alegre}}, \bibinfo {author} {\bibfnamefont {J.}~\bibnamefont {Chan}},
  \bibinfo {author} {\bibfnamefont {M.}~\bibnamefont {Eichenfield}}, \bibinfo
  {author} {\bibfnamefont {M.}~\bibnamefont {Winger}}, \bibinfo {author}
  {\bibfnamefont {Q.}~\bibnamefont {Lin}}, \bibinfo {author} {\bibfnamefont
  {J.~T.}\ \bibnamefont {Hill}}, \bibinfo {author} {\bibfnamefont {D.~E.}\
  \bibnamefont {Chang}}, \ and\ \bibinfo {author} {\bibfnamefont
  {O.}~\bibnamefont {Painter}},\ }\href {\doibase 10.1038/nature09933}
  {\bibfield  {journal} {\bibinfo  {journal} {Nature}\ }\textbf {\bibinfo
  {volume} {472}},\ \bibinfo {pages} {69} (\bibinfo {year} {2011})}\BibitemShut
  {NoStop}%
\bibitem [{\citenamefont {Burgwal}\ and\ \citenamefont
  {Verhagen}(2023)}]{burgwal_enhanced_2023}%
  \BibitemOpen
  \bibfield  {author} {\bibinfo {author} {\bibfnamefont {R.}~\bibnamefont
  {Burgwal}}\ and\ \bibinfo {author} {\bibfnamefont {E.}~\bibnamefont
  {Verhagen}},\ }\href {\doibase 10.1038/s41467-023-37138-z} {\bibfield
  {journal} {\bibinfo  {journal} {Nature Communications}\ }\textbf {\bibinfo
  {volume} {14}},\ \bibinfo {pages} {1526} (\bibinfo {year}
  {2023})}\BibitemShut {NoStop}%
\bibitem [{\citenamefont {Jayich}\ \emph {et~al.}(2008)\citenamefont {Jayich},
  \citenamefont {Sankey}, \citenamefont {Zwickl}, \citenamefont {Yang},
  \citenamefont {Thompson}, \citenamefont {Girvin}, \citenamefont {Clerk},
  \citenamefont {Marquardt},\ and\ \citenamefont
  {Harris}}]{jayich_dispersive_2008}%
  \BibitemOpen
  \bibfield  {author} {\bibinfo {author} {\bibfnamefont {A.~M.}\ \bibnamefont
  {Jayich}}, \bibinfo {author} {\bibfnamefont {J.~C.}\ \bibnamefont {Sankey}},
  \bibinfo {author} {\bibfnamefont {B.~M.}\ \bibnamefont {Zwickl}}, \bibinfo
  {author} {\bibfnamefont {C.}~\bibnamefont {Yang}}, \bibinfo {author}
  {\bibfnamefont {J.~D.}\ \bibnamefont {Thompson}}, \bibinfo {author}
  {\bibfnamefont {S.~M.}\ \bibnamefont {Girvin}}, \bibinfo {author}
  {\bibfnamefont {A.~A.}\ \bibnamefont {Clerk}}, \bibinfo {author}
  {\bibfnamefont {F.}~\bibnamefont {Marquardt}}, \ and\ \bibinfo {author}
  {\bibfnamefont {J.~G.~E.}\ \bibnamefont {Harris}},\ }\href {\doibase
  10.1088/1367-2630/10/9/095008} {\bibfield  {journal} {\bibinfo  {journal}
  {New Journal of Physics}\ }\textbf {\bibinfo {volume} {10}},\ \bibinfo
  {pages} {095008} (\bibinfo {year} {2008})}\BibitemShut {NoStop}%
\bibitem [{\citenamefont {Miri}\ and\ \citenamefont
  {Alù}(2019)}]{miri_exceptional_2019}%
  \BibitemOpen
  \bibfield  {author} {\bibinfo {author} {\bibfnamefont {M.-A.}\ \bibnamefont
  {Miri}}\ and\ \bibinfo {author} {\bibfnamefont {A.}~\bibnamefont {Alù}},\
  }\href {\doibase 10.1126/science.aar7709} {\bibfield  {journal} {\bibinfo
  {journal} {Science}\ }\textbf {\bibinfo {volume} {363}},\ \bibinfo {pages}
  {eaar7709} (\bibinfo {year} {2019})}\BibitemShut {NoStop}%
\bibitem [{\citenamefont {Zhang}\ \emph {et~al.}(2009)\citenamefont {Zhang},
  \citenamefont {Garner},\ and\ \citenamefont {Hau}}]{zhang_creation_2009}%
  \BibitemOpen
  \bibfield  {author} {\bibinfo {author} {\bibfnamefont {R.}~\bibnamefont
  {Zhang}}, \bibinfo {author} {\bibfnamefont {S.~R.}\ \bibnamefont {Garner}}, \
  and\ \bibinfo {author} {\bibfnamefont {L.~V.}\ \bibnamefont {Hau}},\ }\href
  {\doibase 10.1103/PhysRevLett.103.233602} {\bibfield  {journal} {\bibinfo
  {journal} {Physical Review Letters}\ }\textbf {\bibinfo {volume} {103}},\
  \bibinfo {pages} {233602} (\bibinfo {year} {2009})}\BibitemShut {NoStop}%
\bibitem [{\citenamefont {Phillips}\ \emph {et~al.}(2003)\citenamefont
  {Phillips}, \citenamefont {Wang}, \citenamefont {Rumyantsev}, \citenamefont
  {Kwong}, \citenamefont {Takayama},\ and\ \citenamefont
  {Binder}}]{phillips_electromagnetically_2003}%
  \BibitemOpen
  \bibfield  {author} {\bibinfo {author} {\bibfnamefont {M.~C.}\ \bibnamefont
  {Phillips}}, \bibinfo {author} {\bibfnamefont {H.}~\bibnamefont {Wang}},
  \bibinfo {author} {\bibfnamefont {I.}~\bibnamefont {Rumyantsev}}, \bibinfo
  {author} {\bibfnamefont {N.~H.}\ \bibnamefont {Kwong}}, \bibinfo {author}
  {\bibfnamefont {R.}~\bibnamefont {Takayama}}, \ and\ \bibinfo {author}
  {\bibfnamefont {R.}~\bibnamefont {Binder}},\ }\href {\doibase
  10.1103/PhysRevLett.91.183602} {\bibfield  {journal} {\bibinfo  {journal}
  {Physical Review Letters}\ }\textbf {\bibinfo {volume} {91}},\ \bibinfo
  {pages} {183602} (\bibinfo {year} {2003})}\BibitemShut {NoStop}%
\bibitem [{\citenamefont {Wu}\ \emph {et~al.}(2023)\citenamefont {Wu},
  \citenamefont {Cui}, \citenamefont {Xu}, \citenamefont {Feng}, \citenamefont
  {Liu}, \citenamefont {Zhang},\ and\ \citenamefont {Huang}}]{wu_-chip_2023}%
  \BibitemOpen
  \bibfield  {author} {\bibinfo {author} {\bibfnamefont {N.}~\bibnamefont
  {Wu}}, \bibinfo {author} {\bibfnamefont {K.}~\bibnamefont {Cui}}, \bibinfo
  {author} {\bibfnamefont {Q.}~\bibnamefont {Xu}}, \bibinfo {author}
  {\bibfnamefont {X.}~\bibnamefont {Feng}}, \bibinfo {author} {\bibfnamefont
  {F.}~\bibnamefont {Liu}}, \bibinfo {author} {\bibfnamefont {W.}~\bibnamefont
  {Zhang}}, \ and\ \bibinfo {author} {\bibfnamefont {Y.}~\bibnamefont
  {Huang}},\ }\href {\doibase 10.1126/sciadv.abp8892} {\bibfield  {journal}
  {\bibinfo  {journal} {Science Advances}\ }\textbf {\bibinfo {volume} {9}},\
  \bibinfo {pages} {eabp8892} (\bibinfo {year} {2023})}\BibitemShut {NoStop}%
\bibitem [{\citenamefont {Prabhu}\ \emph {et~al.}(2023)\citenamefont {Prabhu},
  \citenamefont {Errando-Herranz}, \citenamefont {De~Santis}, \citenamefont
  {Christen}, \citenamefont {Chen}, \citenamefont {Gerlach},\ and\
  \citenamefont {Englund}}]{prabhu_individually_2023}%
  \BibitemOpen
  \bibfield  {author} {\bibinfo {author} {\bibfnamefont {M.}~\bibnamefont
  {Prabhu}}, \bibinfo {author} {\bibfnamefont {C.}~\bibnamefont
  {Errando-Herranz}}, \bibinfo {author} {\bibfnamefont {L.}~\bibnamefont
  {De~Santis}}, \bibinfo {author} {\bibfnamefont {I.}~\bibnamefont {Christen}},
  \bibinfo {author} {\bibfnamefont {C.}~\bibnamefont {Chen}}, \bibinfo {author}
  {\bibfnamefont {C.}~\bibnamefont {Gerlach}}, \ and\ \bibinfo {author}
  {\bibfnamefont {D.}~\bibnamefont {Englund}},\ }\href {\doibase
  10.1038/s41467-023-37655-x} {\bibfield  {journal} {\bibinfo  {journal}
  {Nature Communications}\ }\textbf {\bibinfo {volume} {14}},\ \bibinfo {pages}
  {2380} (\bibinfo {year} {2023})}\BibitemShut {NoStop}%
\bibitem [{\citenamefont {Shandilya}\ \emph {et~al.}(2021)\citenamefont
  {Shandilya}, \citenamefont {Lake}, \citenamefont {Mitchell}, \citenamefont
  {Sukachev},\ and\ \citenamefont {Barclay}}]{shandilya_optomechanical_2021}%
  \BibitemOpen
  \bibfield  {author} {\bibinfo {author} {\bibfnamefont {P.~K.}\ \bibnamefont
  {Shandilya}}, \bibinfo {author} {\bibfnamefont {D.~P.}\ \bibnamefont {Lake}},
  \bibinfo {author} {\bibfnamefont {M.~J.}\ \bibnamefont {Mitchell}}, \bibinfo
  {author} {\bibfnamefont {D.~D.}\ \bibnamefont {Sukachev}}, \ and\ \bibinfo
  {author} {\bibfnamefont {P.~E.}\ \bibnamefont {Barclay}},\ }\href {\doibase
  10.1038/s41567-021-01364-3} {\bibfield  {journal} {\bibinfo  {journal}
  {Nature Physics}\ }\textbf {\bibinfo {volume} {17}},\ \bibinfo {pages} {1420}
  (\bibinfo {year} {2021})}\BibitemShut {NoStop}%
\bibitem [{\citenamefont {Dumont}\ \emph {et~al.}(2019)\citenamefont {Dumont},
  \citenamefont {Bernard}, \citenamefont {Reinhardt}, \citenamefont {Kato},
  \citenamefont {Ruf},\ and\ \citenamefont
  {Sankey}}]{dumont_flexure-tuned_2019}%
  \BibitemOpen
  \bibfield  {author} {\bibinfo {author} {\bibfnamefont {V.}~\bibnamefont
  {Dumont}}, \bibinfo {author} {\bibfnamefont {S.}~\bibnamefont {Bernard}},
  \bibinfo {author} {\bibfnamefont {C.}~\bibnamefont {Reinhardt}}, \bibinfo
  {author} {\bibfnamefont {A.}~\bibnamefont {Kato}}, \bibinfo {author}
  {\bibfnamefont {M.}~\bibnamefont {Ruf}}, \ and\ \bibinfo {author}
  {\bibfnamefont {J.~C.}\ \bibnamefont {Sankey}},\ }\href {\doibase
  10.1364/OE.27.025731} {\bibfield  {journal} {\bibinfo  {journal} {Optics
  Express}\ }\textbf {\bibinfo {volume} {27}},\ \bibinfo {pages} {25731}
  (\bibinfo {year} {2019})}\BibitemShut {NoStop}%
\bibitem [{\citenamefont {Tagantsev}\ and\ \citenamefont
  {Polzik}(2021)}]{tagantsev_dissipative_2021}%
  \BibitemOpen
  \bibfield  {author} {\bibinfo {author} {\bibfnamefont {A.~K.}\ \bibnamefont
  {Tagantsev}}\ and\ \bibinfo {author} {\bibfnamefont {E.~S.}\ \bibnamefont
  {Polzik}},\ }\href {\doibase 10.1103/PhysRevA.103.063503} {\bibfield
  {journal} {\bibinfo  {journal} {Physical Review A}\ }\textbf {\bibinfo
  {volume} {103}},\ \bibinfo {pages} {063503} (\bibinfo {year}
  {2021})}\BibitemShut {NoStop}%
\bibitem [{\citenamefont {Xuereb}\ \emph {et~al.}(2011)\citenamefont {Xuereb},
  \citenamefont {Schnabel},\ and\ \citenamefont
  {Hammerer}}]{xuereb_dissipative_2011}%
  \BibitemOpen
  \bibfield  {author} {\bibinfo {author} {\bibfnamefont {A.}~\bibnamefont
  {Xuereb}}, \bibinfo {author} {\bibfnamefont {R.}~\bibnamefont {Schnabel}}, \
  and\ \bibinfo {author} {\bibfnamefont {K.}~\bibnamefont {Hammerer}},\ }\href
  {\doibase 10.1103/PhysRevLett.107.213604} {\bibfield  {journal} {\bibinfo
  {journal} {Physical Review Letters}\ }\textbf {\bibinfo {volume} {107}},\
  \bibinfo {pages} {213604} (\bibinfo {year} {2011})}\BibitemShut {NoStop}%
\bibitem [{\citenamefont {Kyriienko}\ \emph {et~al.}(2014)\citenamefont
  {Kyriienko}, \citenamefont {Liew},\ and\ \citenamefont
  {Shelykh}}]{kyriienko_optomechanics_2014}%
  \BibitemOpen
  \bibfield  {author} {\bibinfo {author} {\bibfnamefont {O.}~\bibnamefont
  {Kyriienko}}, \bibinfo {author} {\bibfnamefont {T.~C.~H.}\ \bibnamefont
  {Liew}}, \ and\ \bibinfo {author} {\bibfnamefont {I.~A.}\ \bibnamefont
  {Shelykh}},\ }\href {\doibase 10.1103/PhysRevLett.112.076402} {\bibfield
  {journal} {\bibinfo  {journal} {Physical Review Letters}\ }\textbf {\bibinfo
  {volume} {112}},\ \bibinfo {pages} {076402} (\bibinfo {year}
  {2014})}\BibitemShut {NoStop}%
\bibitem [{\citenamefont {Primo}\ \emph {et~al.}(2021)\citenamefont {Primo},
  \citenamefont {Kersul}, \citenamefont {Benevides}, \citenamefont {Carvalho},
  \citenamefont {Ménard}, \citenamefont {Frateschi}, \citenamefont {de~Assis},
  \citenamefont {Wiederhecker},\ and\ \citenamefont
  {Mayer~Alegre}}]{primo_accurate_2021}%
  \BibitemOpen
  \bibfield  {author} {\bibinfo {author} {\bibfnamefont {A.~G.}\ \bibnamefont
  {Primo}}, \bibinfo {author} {\bibfnamefont {C.~M.}\ \bibnamefont {Kersul}},
  \bibinfo {author} {\bibfnamefont {R.}~\bibnamefont {Benevides}}, \bibinfo
  {author} {\bibfnamefont {N.~C.}\ \bibnamefont {Carvalho}}, \bibinfo {author}
  {\bibfnamefont {M.}~\bibnamefont {Ménard}}, \bibinfo {author} {\bibfnamefont
  {N.~C.}\ \bibnamefont {Frateschi}}, \bibinfo {author} {\bibfnamefont {P.-L.}\
  \bibnamefont {de~Assis}}, \bibinfo {author} {\bibfnamefont {G.~S.}\
  \bibnamefont {Wiederhecker}}, \ and\ \bibinfo {author} {\bibfnamefont
  {T.~P.}\ \bibnamefont {Mayer~Alegre}},\ }\href {\doibase 10.1063/5.0055201}
  {\bibfield  {journal} {\bibinfo  {journal} {APL Photonics}\ }\textbf
  {\bibinfo {volume} {6}},\ \bibinfo {pages} {086101} (\bibinfo {year}
  {2021})}\BibitemShut {NoStop}%
\bibitem [{\citenamefont {Barclay}\ \emph {et~al.}(2005)\citenamefont
  {Barclay}, \citenamefont {Srinivasan},\ and\ \citenamefont
  {Painter}}]{barclay_nonlinear_2005}%
  \BibitemOpen
  \bibfield  {author} {\bibinfo {author} {\bibfnamefont {P.~E.}\ \bibnamefont
  {Barclay}}, \bibinfo {author} {\bibfnamefont {K.}~\bibnamefont {Srinivasan}},
  \ and\ \bibinfo {author} {\bibfnamefont {O.}~\bibnamefont {Painter}},\ }\href
  {\doibase 10.1364/OPEX.13.000801} {\bibfield  {journal} {\bibinfo  {journal}
  {Optics Express}\ }\textbf {\bibinfo {volume} {13}},\ \bibinfo {pages} {801}
  (\bibinfo {year} {2005})}\BibitemShut {NoStop}%
\bibitem [{\citenamefont {Primo}\ \emph {et~al.}(2023)\citenamefont {Primo},
  \citenamefont {Pinho}, \citenamefont {Benevides}, \citenamefont
  {Gr\"oblacher}, \citenamefont {Wiederhecker},\ and\ \citenamefont
  {Alegre}}]{zenodo_data}%
  \BibitemOpen
  \bibfield  {author} {\bibinfo {author} {\bibfnamefont {A.~G.}\ \bibnamefont
  {Primo}}, \bibinfo {author} {\bibfnamefont {P.~V.}\ \bibnamefont {Pinho}},
  \bibinfo {author} {\bibfnamefont {R.}~\bibnamefont {Benevides}}, \bibinfo
  {author} {\bibfnamefont {S.}~\bibnamefont {Gr\"oblacher}}, \bibinfo {author}
  {\bibfnamefont {G.~S.}\ \bibnamefont {Wiederhecker}}, \ and\ \bibinfo
  {author} {\bibfnamefont {T.~P.~M.}\ \bibnamefont {Alegre}},\ }\href {\doibase
  10.5281/zenodo.8072537} {\enquote {\bibinfo {title} {{Data and simulation
  files for: ``Dissipative Optomechanics in High-Frequency Nanomechanical
  Resonators"}},}\ } (\bibinfo {year} {2023})\BibitemShut {NoStop}%
\end{thebibliography}%


\begin{thebibliography}{14}%
\makeatletter
\providecommand \@ifxundefined [1]{%
 \@ifx{#1\undefined}
}%
\providecommand \@ifnum [1]{%
 \ifnum #1\expandafter \@firstoftwo
 \else \expandafter \@secondoftwo
 \fi
}%
\providecommand \@ifx [1]{%
 \ifx #1\expandafter \@firstoftwo
 \else \expandafter \@secondoftwo
 \fi
}%
\providecommand \natexlab [1]{#1}%
\providecommand \enquote  [1]{``#1''}%
\providecommand \bibnamefont  [1]{#1}%
\providecommand \bibfnamefont [1]{#1}%
\providecommand \citenamefont [1]{#1}%
\providecommand \href@noop [0]{\@secondoftwo}%
\providecommand \href [0]{\begingroup \@sanitize@url \@href}%
\providecommand \@href[1]{\@@startlink{#1}\@@href}%
\providecommand \@@href[1]{\endgroup#1\@@endlink}%
\providecommand \@sanitize@url [0]{\catcode `\\12\catcode `\$12\catcode
  `\&12\catcode `\#12\catcode `\^12\catcode `\_12\catcode `\%12\relax}%
\providecommand \@@startlink[1]{}%
\providecommand \@@endlink[0]{}%
\providecommand \url  [0]{\begingroup\@sanitize@url \@url }%
\providecommand \@url [1]{\endgroup\@href {#1}{\urlprefix }}%
\providecommand \urlprefix  [0]{URL }%
\providecommand \Eprint [0]{\href }%
\providecommand \doibase [0]{http://dx.doi.org/}%
\providecommand \selectlanguage [0]{\@gobble}%
\providecommand \bibinfo  [0]{\@secondoftwo}%
\providecommand \bibfield  [0]{\@secondoftwo}%
\providecommand \translation [1]{[#1]}%
\providecommand \BibitemOpen [0]{}%
\providecommand \bibitemStop [0]{}%
\providecommand \bibitemNoStop [0]{.\EOS\space}%
\providecommand \EOS [0]{\spacefactor3000\relax}%
\providecommand \BibitemShut  [1]{\csname bibitem#1\endcsname}%
\let\auto@bib@innerbib\@empty
\bibitem [{\citenamefont {Haus}(1984)}]{Haus1984WavesOptoelectronics}%
  \BibitemOpen
  \bibfield  {author} {\bibinfo {author} {\bibfnamefont {H.}~\bibnamefont
  {Haus}},\ }\href {https://books.google.com.br/books?id=AQUpAQAAMAAJ} {\emph
  {\bibinfo {title} {{Waves and Fields in Optoelectronics}}}}\ (\bibinfo
  {publisher} {Prentice-Hall},\ \bibinfo {year} {1984})\BibitemShut {NoStop}%
\bibitem [{\citenamefont {Yanay}\ \emph {et~al.}(2016)\citenamefont {Yanay},
  \citenamefont {Sankey},\ and\ \citenamefont
  {Clerk}}]{Yanay2016QuantumSystems}%
  \BibitemOpen
  \bibfield  {author} {\bibinfo {author} {\bibfnamefont {Y.}~\bibnamefont
  {Yanay}}, \bibinfo {author} {\bibfnamefont {J.~C.}\ \bibnamefont {Sankey}}, \
  and\ \bibinfo {author} {\bibfnamefont {A.~A.}\ \bibnamefont {Clerk}},\ }\href
  {\doibase 10.1103/PHYSREVA.93.063809} {\bibfield  {journal} {\bibinfo
  {journal} {Physical Review A}\ }\textbf {\bibinfo {volume} {93}},\ \bibinfo
  {pages} {063809} (\bibinfo {year} {2016})}\BibitemShut {NoStop}%
\bibitem [{\citenamefont {Elste}\ \emph {et~al.}(2009)\citenamefont {Elste},
  \citenamefont {Girvin},\ and\ \citenamefont
  {Clerk}}]{Elste2009QuantumNanomechanics}%
  \BibitemOpen
  \bibfield  {author} {\bibinfo {author} {\bibfnamefont {F.}~\bibnamefont
  {Elste}}, \bibinfo {author} {\bibfnamefont {S.~M.}\ \bibnamefont {Girvin}}, \
  and\ \bibinfo {author} {\bibfnamefont {A.~A.}\ \bibnamefont {Clerk}},\ }\href
  {\doibase 10.1103/PhysRevLett.102.207209} {\bibfield  {journal} {\bibinfo
  {journal} {Physical Review Letters}\ }\textbf {\bibinfo {volume} {102}},\
  \bibinfo {pages} {207209} (\bibinfo {year} {2009})}\BibitemShut {NoStop}%
\bibitem [{\citenamefont {Weiss}\ \emph {et~al.}(2013)\citenamefont {Weiss},
  \citenamefont {Bruder},\ and\ \citenamefont
  {Nunnenkamp}}]{Weiss2013Strong-couplingSystems}%
  \BibitemOpen
  \bibfield  {author} {\bibinfo {author} {\bibfnamefont {T.}~\bibnamefont
  {Weiss}}, \bibinfo {author} {\bibfnamefont {C.}~\bibnamefont {Bruder}}, \
  and\ \bibinfo {author} {\bibfnamefont {A.}~\bibnamefont {Nunnenkamp}},\
  }\href {\doibase 10.1088/1367-2630/15/4/045017} {\bibfield  {journal}
  {\bibinfo  {journal} {New Journal of Physics}\ }\textbf {\bibinfo {volume}
  {15}},\ \bibinfo {pages} {045017} (\bibinfo {year} {2013})}\BibitemShut
  {NoStop}%
\bibitem [{\citenamefont {Kubo}(1966)}]{Kubo1966TheTheorem}%
  \BibitemOpen
  \bibfield  {author} {\bibinfo {author} {\bibfnamefont {R.}~\bibnamefont
  {Kubo}},\ }\href {\doibase 10.1088/0034-4885/29/1/306} {\bibfield  {journal}
  {\bibinfo  {journal} {Reports on Progress in Physics}\ }\textbf {\bibinfo
  {volume} {29}},\ \bibinfo {pages} {255} (\bibinfo {year} {1966})}\BibitemShut
  {NoStop}%
\bibitem [{\citenamefont {Primo}\ \emph {et~al.}(2020)\citenamefont {Primo},
  \citenamefont {Carvalho}, \citenamefont {Kersul}, \citenamefont {Frateschi},
  \citenamefont {Wiederhecker},\ and\ \citenamefont
  {Alegre}}]{Primo2020Quasinormal-ModeOptomechanics}%
  \BibitemOpen
  \bibfield  {author} {\bibinfo {author} {\bibfnamefont {A.~G.}\ \bibnamefont
  {Primo}}, \bibinfo {author} {\bibfnamefont {N.~C.}\ \bibnamefont {Carvalho}},
  \bibinfo {author} {\bibfnamefont {C.~M.}\ \bibnamefont {Kersul}}, \bibinfo
  {author} {\bibfnamefont {N.~C.}\ \bibnamefont {Frateschi}}, \bibinfo {author}
  {\bibfnamefont {G.~S.}\ \bibnamefont {Wiederhecker}}, \ and\ \bibinfo
  {author} {\bibfnamefont {T.~P.}\ \bibnamefont {Alegre}},\ }\href {\doibase
  10.1103/PHYSREVLETT.125.233601} {\bibfield  {journal} {\bibinfo  {journal}
  {Physical Review Letters}\ }\textbf {\bibinfo {volume} {125}},\ \bibinfo
  {pages} {233601} (\bibinfo {year} {2020})}\BibitemShut {NoStop}%
\bibitem [{\citenamefont {Safavi-Naeini}\ \emph {et~al.}(2011)\citenamefont
  {Safavi-Naeini}, \citenamefont {Alegre}, \citenamefont {Chan}, \citenamefont
  {Eichenfield}, \citenamefont {Winger}, \citenamefont {Lin}, \citenamefont
  {Hill}, \citenamefont {Chang},\ and\ \citenamefont
  {Painter}}]{Safavi-Naeini2011ElectromagneticallyOptomechanics}%
  \BibitemOpen
  \bibfield  {author} {\bibinfo {author} {\bibfnamefont {A.~H.}\ \bibnamefont
  {Safavi-Naeini}}, \bibinfo {author} {\bibfnamefont {T.~P.}\ \bibnamefont
  {Alegre}}, \bibinfo {author} {\bibfnamefont {J.}~\bibnamefont {Chan}},
  \bibinfo {author} {\bibfnamefont {M.}~\bibnamefont {Eichenfield}}, \bibinfo
  {author} {\bibfnamefont {M.}~\bibnamefont {Winger}}, \bibinfo {author}
  {\bibfnamefont {Q.}~\bibnamefont {Lin}}, \bibinfo {author} {\bibfnamefont
  {J.~T.}\ \bibnamefont {Hill}}, \bibinfo {author} {\bibfnamefont {D.~E.}\
  \bibnamefont {Chang}}, \ and\ \bibinfo {author} {\bibfnamefont
  {O.}~\bibnamefont {Painter}},\ }\href {\doibase 10.1038/nature09933}
  {\bibfield  {journal} {\bibinfo  {journal} {Nature 2011 472:7341}\ }\textbf
  {\bibinfo {volume} {472}},\ \bibinfo {pages} {69} (\bibinfo {year}
  {2011})}\BibitemShut {NoStop}%
\bibitem [{\citenamefont {Weis}\ \emph {et~al.}(2010)\citenamefont {Weis},
  \citenamefont {Rivi{\`{e}}re}, \citenamefont {Del{\'{e}}glise}, \citenamefont
  {Gavartin}, \citenamefont {Arcizet}, \citenamefont {Schliesser},\ and\
  \citenamefont {Kippenberg}}]{Weis2010OptomechanicallyTransparency}%
  \BibitemOpen
  \bibfield  {author} {\bibinfo {author} {\bibfnamefont {S.}~\bibnamefont
  {Weis}}, \bibinfo {author} {\bibfnamefont {R.}~\bibnamefont {Rivi{\`{e}}re}},
  \bibinfo {author} {\bibfnamefont {S.}~\bibnamefont {Del{\'{e}}glise}},
  \bibinfo {author} {\bibfnamefont {E.}~\bibnamefont {Gavartin}}, \bibinfo
  {author} {\bibfnamefont {O.}~\bibnamefont {Arcizet}}, \bibinfo {author}
  {\bibfnamefont {A.}~\bibnamefont {Schliesser}}, \ and\ \bibinfo {author}
  {\bibfnamefont {T.~J.}\ \bibnamefont {Kippenberg}},\ }\href {\doibase
  10.1126/science.1195596} {\bibfield  {journal} {\bibinfo  {journal}
  {Science}\ }\textbf {\bibinfo {volume} {330}},\ \bibinfo {pages} {1520}
  (\bibinfo {year} {2010})}\BibitemShut {NoStop}%
\bibitem [{\citenamefont {Shomroni}\ \emph {et~al.}(2019)\citenamefont
  {Shomroni}, \citenamefont {Qiu}, \citenamefont {Malz}, \citenamefont
  {Nunnenkamp},\ and\ \citenamefont
  {Kippenberg}}]{Shomroni2019OpticalOscillator}%
  \BibitemOpen
  \bibfield  {author} {\bibinfo {author} {\bibfnamefont {I.}~\bibnamefont
  {Shomroni}}, \bibinfo {author} {\bibfnamefont {L.}~\bibnamefont {Qiu}},
  \bibinfo {author} {\bibfnamefont {D.}~\bibnamefont {Malz}}, \bibinfo {author}
  {\bibfnamefont {A.}~\bibnamefont {Nunnenkamp}}, \ and\ \bibinfo {author}
  {\bibfnamefont {T.~J.}\ \bibnamefont {Kippenberg}},\ }\href {\doibase
  10.1038/s41467-019-10024-3} {\bibfield  {journal} {\bibinfo  {journal}
  {Nature Communications 2019 10:1}\ }\textbf {\bibinfo {volume} {10}},\
  \bibinfo {pages} {1} (\bibinfo {year} {2019})}\BibitemShut {NoStop}%
\bibitem [{\citenamefont {Primo}\ \emph {et~al.}(2021)\citenamefont {Primo},
  \citenamefont {Kersul}, \citenamefont {Benevides}, \citenamefont {Carvalho},
  \citenamefont {M{\'{e}}nard}, \citenamefont {Frateschi}, \citenamefont
  {De~Assis}, \citenamefont {Wiederhecker},\ and\ \citenamefont
  {Mayer~Alegre}}]{Primo2021AccurateOptomechanics}%
  \BibitemOpen
  \bibfield  {author} {\bibinfo {author} {\bibfnamefont {A.~G.}\ \bibnamefont
  {Primo}}, \bibinfo {author} {\bibfnamefont {C.~M.}\ \bibnamefont {Kersul}},
  \bibinfo {author} {\bibfnamefont {R.}~\bibnamefont {Benevides}}, \bibinfo
  {author} {\bibfnamefont {N.~C.}\ \bibnamefont {Carvalho}}, \bibinfo {author}
  {\bibfnamefont {M.}~\bibnamefont {M{\'{e}}nard}}, \bibinfo {author}
  {\bibfnamefont {N.~C.}\ \bibnamefont {Frateschi}}, \bibinfo {author}
  {\bibfnamefont {P.~L.}\ \bibnamefont {De~Assis}}, \bibinfo {author}
  {\bibfnamefont {G.~S.}\ \bibnamefont {Wiederhecker}}, \ and\ \bibinfo
  {author} {\bibfnamefont {T.~P.}\ \bibnamefont {Mayer~Alegre}},\ }\href
  {\doibase 10.1063/5.0055201} {\bibfield  {journal} {\bibinfo  {journal} {APL
  Photonics}\ }\textbf {\bibinfo {volume} {6}},\ \bibinfo {pages} {086101}
  (\bibinfo {year} {2021})}\BibitemShut {NoStop}%
\bibitem [{\citenamefont {Barclay}\ \emph {et~al.}(2005)\citenamefont
  {Barclay}, \citenamefont {Srinivasan},\ and\ \citenamefont
  {Painter}}]{Barclay2005NonlinearTaper}%
  \BibitemOpen
  \bibfield  {author} {\bibinfo {author} {\bibfnamefont {P.~E.}\ \bibnamefont
  {Barclay}}, \bibinfo {author} {\bibfnamefont {K.}~\bibnamefont {Srinivasan}},
  \ and\ \bibinfo {author} {\bibfnamefont {O.}~\bibnamefont {Painter}},\ }\href
  {\doibase 10.1364/OPEX.13.000801} {\bibfield  {journal} {\bibinfo  {journal}
  {Optics Express, Vol. 13, Issue 3, pp. 801-820}\ }\textbf {\bibinfo {volume}
  {13}},\ \bibinfo {pages} {801} (\bibinfo {year} {2005})}\BibitemShut
  {NoStop}%
\bibitem [{\citenamefont {Wallucks}\ \emph {et~al.}(2020)\citenamefont
  {Wallucks}, \citenamefont {Marinkovi{\'{c}}}, \citenamefont {Hensen},
  \citenamefont {Stockill},\ and\ \citenamefont
  {Gr{\"{o}}blacher}}]{Wallucks2020AWavelengths}%
  \BibitemOpen
  \bibfield  {author} {\bibinfo {author} {\bibfnamefont {A.}~\bibnamefont
  {Wallucks}}, \bibinfo {author} {\bibfnamefont {I.}~\bibnamefont
  {Marinkovi{\'{c}}}}, \bibinfo {author} {\bibfnamefont {B.}~\bibnamefont
  {Hensen}}, \bibinfo {author} {\bibfnamefont {R.}~\bibnamefont {Stockill}}, \
  and\ \bibinfo {author} {\bibfnamefont {S.}~\bibnamefont {Gr{\"{o}}blacher}},\
  }\href {\doibase 10.1038/s41567-020-0891-z} {\bibfield  {journal} {\bibinfo
  {journal} {Nature Physics 2020 16:7}\ }\textbf {\bibinfo {volume} {16}},\
  \bibinfo {pages} {772} (\bibinfo {year} {2020})}\BibitemShut {NoStop}%
\bibitem [{\citenamefont {Jiang}\ \emph {et~al.}(2022)\citenamefont {Jiang},
  \citenamefont {Mayor}, \citenamefont {Malik}, \citenamefont {Van~Laer},
  \citenamefont {McKenna}, \citenamefont {Patel}, \citenamefont {Witmer},\ and\
  \citenamefont {Safavi-Naeini}}]{Jiang2022OpticallyPhotons}%
  \BibitemOpen
  \bibfield  {author} {\bibinfo {author} {\bibfnamefont {W.}~\bibnamefont
  {Jiang}}, \bibinfo {author} {\bibfnamefont {F.~M.}\ \bibnamefont {Mayor}},
  \bibinfo {author} {\bibfnamefont {S.}~\bibnamefont {Malik}}, \bibinfo
  {author} {\bibfnamefont {R.}~\bibnamefont {Van~Laer}}, \bibinfo {author}
  {\bibfnamefont {T.~P.}\ \bibnamefont {McKenna}}, \bibinfo {author}
  {\bibfnamefont {R.~N.}\ \bibnamefont {Patel}}, \bibinfo {author}
  {\bibfnamefont {J.~D.}\ \bibnamefont {Witmer}}, \ and\ \bibinfo {author}
  {\bibfnamefont {A.~H.}\ \bibnamefont {Safavi-Naeini}},\ }\href {\doibase
  10.48550/arxiv.2210.10739} {\  (\bibinfo {year} {2022}),\
  10.48550/arxiv.2210.10739}\BibitemShut {NoStop}%
\bibitem [{\citenamefont {Primo}\ \emph {et~al.}(2023)\citenamefont {Primo},
  \citenamefont {Pinho}, \citenamefont {Benevides}, \citenamefont
  {Gr\"oblacher}, \citenamefont {Wiederhecker},\ and\ \citenamefont
  {Alegre}}]{zenodo_data}%
  \BibitemOpen
  \bibfield  {author} {\bibinfo {author} {\bibfnamefont {A.~G.}\ \bibnamefont
  {Primo}}, \bibinfo {author} {\bibfnamefont {P.~V.}\ \bibnamefont {Pinho}},
  \bibinfo {author} {\bibfnamefont {R.}~\bibnamefont {Benevides}}, \bibinfo
  {author} {\bibfnamefont {S.}~\bibnamefont {Gr\"oblacher}}, \bibinfo {author}
  {\bibfnamefont {G.~S.}\ \bibnamefont {Wiederhecker}}, \ and\ \bibinfo
  {author} {\bibfnamefont {T.~P.~M.}\ \bibnamefont {Alegre}},\ }\href {\doibase
  10.5281/zenodo.8072537} {\enquote {\bibinfo {title} {{Data and simulation
  files for: ``Dissipative Optomechanics in High-Frequency Nanomechanical
  Resonators"}},}\ } (\bibinfo {year} {2023})\BibitemShut {NoStop}%
\end{thebibliography}%

%

\end{document}


\title{Supplemental Material: Dissipative Optomechanics in High-Frequency Nanomechanical Resonators}
\author{Andr\'e G. Primo}
\email{agprimo@ifi.unicamp.br, ppinho@ifi.unicamp.br}
\thanks{These authors contributed equally to this work.}
\affiliation{Gleb Wataghin Institute of Physics, University of Campinas, 13083-859 Campinas, SP, Brazil}

\author{Pedro V. Pinho}
\email{agprimo@ifi.unicamp.br, ppinho@ifi.unicamp.br}
\thanks{These authors contributed equally to this work.}
\affiliation{Gleb Wataghin Institute of Physics, University of Campinas, 13083-859 Campinas, SP, Brazil}

\author{Rodrigo Benevides}
\affiliation{Department of Physics, ETH Zürich, 8093 Zürich, Switzerland}

\author{Simon Gr\"oblacher}
\affiliation{Kavli Institute of Nanoscience, Department of Quantum Nanoscience, Delft University of Technology, Lorentzweg 1, 2628CJ Delft, The Netherlands}

\author{Gustavo S. Wiederhecker}
\affiliation{Gleb Wataghin Institute of Physics, University of Campinas, 13083-859 Campinas, SP, Brazil}

\author{Thiago P. Mayer Alegre}
\email{alegre@unicamp.br}
\affiliation{Gleb Wataghin Institute of Physics, University of Campinas, 13083-859 Campinas, SP, Brazil}

\date{\today}

\maketitle

\section{From an eigenvalue problem to dissipative optomechanics}\label{S1}

Here we aim at deriving the existence of a dissipative optomechanical coupling using the system of coupled cavities in Fig.\textbf{1b} of the main text. We generalize the discussion by allowing both ``bare" optical modes to display different frequencies $\omega_1$ and $\omega_2 =  \omega_1 - \Delta\omega$. The amplitude of the optical fields $a_1$ and $a_2$ are described as\cite{Haus1984WavesOptoelectronics}:

\begin{equation}
   \frac{d}{dt} \begin{bmatrix}
a_1\\
a_2
\end{bmatrix} = \begin{bmatrix}
-i \left(\omega_1- G_\omega x\right) - \frac{(\kappa_e+\kappa_i)}{2} & iJ\\
iJ & -i\left( \omega_1-\Delta\omega \right) - \frac{\kappa_i}{2}
\end{bmatrix}\cdot
\begin{bmatrix}
a_1\\
a_2
\end{bmatrix}
- \begin{bmatrix}
\sqrt{\kappa_e} \bar{\alpha}_\text{in} e^{-i \omega_l t}\\
0
\end{bmatrix},
\label{eq:eigv}
\end{equation}
which can be written in the form of $\frac{d}{dt}\vec{a} = \mathbf{M}\cdot \vec{a} + \mathbf{J}$ and $\bar{\alpha}_\text{in}$ is a coherent driving amplitude. 

In the experiment, we probe this system through its supermode response, or the eigenvectors of the matrix $\mathbf{M}$. The eigenvalues of $\mathbf{M}$, $E_\pm$, are given by:

\begin{equation}
    E_\mp = -i\left(\omega_1 - \frac{\Delta\omega}{2} - \frac{G_\omega}{2}x\right)-\frac{\kappa_i}{2} -  \frac{\kappa_e}{4} \pm i \sqrt{J^2 + \left[\frac{2(-\Delta\omega + G_\omega x) + i \kappa_e}{4}\right]^2},
\end{equation}
which can be Taylor expanded keeping terms only to the first order in $x$, yielding:

\begin{equation}
    E_\mp =  -i\left(\omega_1 - \frac{\Delta\omega}{2}\right)-\frac{\kappa_i}{2} -  \frac{\kappa_e}{4} \pm i \sqrt{J^2 + \left(\frac{-2\Delta\omega + i \kappa_e}{4}\right)^2} + \left(i\frac{G_\omega}{2} \mp \frac{G_\omega}{2} \frac{i 2\Delta\omega + \kappa_e}{4\sqrt{J^2 + \left(\frac{2\Delta\omega - i \kappa_e}{4}\right)^2}}\right) x.
\end{equation}

The terms independent of $x$ define the unperturbed response of the coupled resonators. Their imaginary (real) parts correspond to the frequencies (loss rates) of the supermodes. In the notation adopted in the main text, $E_+$ is associated with $\omega_+$, $\kappa_+$ and $E_-$ with $\omega_-$, $\kappa_-$. In the regime $J \gg \kappa_e$, corresponding to our experiment, the eigenfrequencies split into two branches. This is shown in Fig.~1\textbf{c} of the main text, as a function of $\Delta\omega$. Naturally,  if $|\Delta\omega| \gg J$ we should recover the uncoupled cavity limit. This is mathematically equivalent to taking $J \to 0$ in the equations above. Importantly, our device operates in a regime $\Delta\omega/J \ll 1$, which is near optimal. In fact, for $\Delta\omega = 0$ we note that the losses of both supermodes are identical and their frequencies differ by a factor $2\sqrt{J^2-\kappa_e^2/16} \approx 2J$, indicating a perfect optical hybridization, that is, $\omega_\pm \approx  \omega_1 \pm J$ and $\kappa_\pm = \kappa_i + \frac{\kappa_e}{2}$.  From now on, we assume $\Delta\omega \to 0$.

The linear terms on $x$ generate the optomechanical coupling. For simplicity, we may consider the terms multiplying $x$ as an effective optomechanical coupling. We have

\begin{equation}
     \lim_{\Delta\omega \to 0}\left(i\frac{G_\omega}{2} \mp \frac{G_\omega}{2} \frac{i 2\Delta\omega + \kappa_e}{4\sqrt{J^2 + \left(\frac{2\Delta\omega - i \kappa_e}{4}\right)^2}}\right) x =  \left(i\frac{G_\omega}{2} \mp \frac{G_\omega}{2} \frac{\kappa_e}{4\sqrt{J^2 - \left(\frac{\kappa_e}{4}\right)^2}}\right) x,
\end{equation}
from which we see that the effective optomechanical coupling has both an imaginary and a real part, which are associated with dispersive and dissipative couplings, respectively~\cite{Yanay2016QuantumSystems}. Here, the dispersive component is divided by a factor of $2$ when compared to the uncoupled cavity regime, and is identical for both supermodes, i.e. $G_{\omega_\pm} = \frac{G_\omega}{2}$. The effective dissipative coupling is given by
\begin{equation}
    G_{\kappa_{e_\pm}} =  \mp G_\omega \frac{\kappa_e}{4\sqrt{J^2-(\frac{\kappa_e}{4})^2}},
    \label{eq:gk_theory}
\end{equation}
and associated with a modulation in the extrinsic losses, since this is the dominant loss asymmetry in our analysis. In our experiment, $g_\omega/(2\pi) \approx \SI{700}{\kHz}$, $\kappa_e/(2\pi) \approx \SI{280}{\MHz}$, and $J/(2\pi)\approx \SI{15}{\GHz}$, where $g_\omega$ and $\kappa_e$ were obtained by summing the measured values of $g_{\omega_\pm}$ and $\kappa_{e_\pm}$. This is accurate since the eigenvalue treatment of the problem points out that the bare cavity dispersive coupling and extrinsic loss are, in fact, split between the two supermodes (if $\Delta\omega \neq 0$ this splitting is uneven). Plugging the experimental values in Eq.~\ref{eq:gk_theory}, we find $g_{\kappa_{e_\pm}}/(2\pi) \approx \mp \SI{3.26}{\kHz}$, in excellent agreement with our measurements.

The source term in Eq.~\ref{eq:eigv} is treated as follows: the eigenvectors of $\mathbf{M}$, $a_\pm$,  are related to $a_1$ and $a_2$ through:

\begin{equation}
     \begin{bmatrix}
a_+\\
a_-
\end{bmatrix} =  \mathbf{P}(x) \cdot \begin{bmatrix}
a_1\\
a_2
\end{bmatrix},
\end{equation}
where $\mathbf{P}(x)$ is the basis transformation matrix. If $\Omega \ll 2J$, i.e. the adiabatic regime, we may assume that the time variations of $\mathbf{P}(x)$ are slow compared to those of $a_1$ and $a_2$. This can be readily seen by moving into a frame rotating at the laser frequency $\omega_l$. In this case, we can rewrite Eq.~\ref{eq:eigv} in terms of the column vector $\vec{a}_S =  [a_+, a_-]^t$ as:

\begin{equation}
    \frac{d}{dt} \vec{a}_S = \mathbf{P} \cdot \mathbf{M} \cdot \mathbf{P}^{-1} \cdot \vec{a}_S + \mathbf{P} \cdot \mathbf{J},
\end{equation}
meaning the driving term is now dependent on $x$. The matrix $\mathbf{P} \cdot \mathbf{M} \cdot \mathbf{P}^{-1}$ is diagonal with the eigenvalues $E_\pm$ as its non-zero terms. The matrix $\mathbf{P}$ is given by:

\begin{equation}
    \mathbf{P}(x) =  \frac{1}{\sqrt{2}}
    \renewcommand*{\arraystretch}{2}
\begin{bmatrix}
-1- \frac{G_{\kappa_{e_+}}}{\kappa_e}x & 1 - \frac{G_{\kappa_{e_+}}}{\kappa_e}x\\
 1 + \frac{G_{\kappa_{e_-}}}{\kappa_e}x & 1 - \frac{G_{\kappa_{e_-}}}{\kappa_e}x
\end{bmatrix}
,
\end{equation}
where we also used $\kappa_e \ll J$ and neglected nonlinear terms in $x$. Finally, defining $\kappa_{e_\pm} =  \frac{\kappa_e}{2}$ we have:

\begin{equation}
    \frac{d}{dt} \begin{bmatrix}
a_+\\
a_-
\end{bmatrix} = 
\setlength\arraycolsep{-10pt}
\begin{bmatrix}
-i \left(\omega_+ - G_{\omega_+} x\right) - \frac{(\kappa_+ + G_{\kappa_{e_+}}x)}{2} & 0\\
0 & -i \left(\omega_- - G_{\omega_-} x\right) - \frac{(\kappa_- + G_{\kappa_{e_-}}x)}{2}
\end{bmatrix}\setlength\arraycolsep{0pt}\cdot
\begin{bmatrix}
a_+\\
a_-
\end{bmatrix}
-\begingroup
\renewcommand*{\arraystretch}{2.5}
\begin{bmatrix}
-\sqrt{\kappa_{e_+}}-\frac{G_{\kappa_{e_+}}}{2 \sqrt{\kappa_{e_+}}}x\\
\sqrt{\kappa_{e_-}}+\frac{G_{\kappa_{e_-}}}{2 \sqrt{\kappa_{e_-}}}x
\end{bmatrix}\endgroup
 \bar{\alpha}_\text{in} e^{-i \omega_l t}.
\end{equation}
\section{Generalized optomechanical transduction}

We turn our attention to the description of the fluctuations imprinted in the optical field due to the thermo-mechanical noise. Since our supermodes are several mechanical frequencies away from each other, we can analyze their response individually.  We start from the classical optical field description:

\begin{equation}
\dot a = i (\Delta + G_\omega x )a  -\frac{\kappa + G_{\kappa_e}x }{2}a -\sqrt{\kappa_e} \bar{\alpha}_\text{in} - \frac{G_{\kappa_e}x}{2\sqrt{\kappa_e}}  \bar{\alpha}_\text{in},
\label{eq:rate_a}
\end{equation}
where we moved into a reference frame rotating at the laser frequency $\omega_l$, therefore, $\Delta =  \omega_l - \omega_c$, where $\omega_c$ is the mode's frequency. Here, we can associate $a$, $\Delta$, $\kappa$, and $\kappa_e$ with either of the supermodes. We can linearize this equation around stationary coherent amplitudes $\bar{x}$ and $\bar{a}$ by making $a(t) = \bar{a} + \delta a (t)$ and $x(t) = \bar{x} +  \delta x (t)$, and keeping terms only to first order in the fluctuations. The coherent electromagnetic field is given by:

 \begin{equation}
     \bar{a} = \frac{ \sqrt{\bar{\kappa}_e}\bar{\alpha}_\text{in}}{i\bar\Delta - \frac{\bar{\kappa}}{2}},
 \end{equation}
 and the dynamical part reads:

 \begin{equation}
      \dot{\delta a} = i (\bar \Delta  \delta a + G_\omega \bar{a} \delta x )  -\frac{\bar\kappa \delta a + G_{\kappa_e}\bar{a} \delta x }{2} - \frac{G_{\kappa_e} \delta x}{2\sqrt{\bar{\kappa}_e}}  \bar{\alpha}_\text{in},
      \label{eq:dyn}
 \end{equation}
 where $\bar\Delta$ , $\bar \kappa$, $\bar{\kappa}_e$ are the renormalized detuning, total and extrinsic losses, respectively, accounting for the effects arising from a static shift in the mechanical displacement, $\bar{x}$. 
 
 The form of Eq.~\ref{eq:dyn} allows a clear interpretation of the several sources contributing to the optical field fluctuations. First, we have the dispersively scattered part of $\delta a$, which is seeded by the pump photons $\bar{a}$. On the dissipative side, we have source terms proportional to $\bar a$ and to $\bar{\alpha}_\text{in}$. The first is enhanced by the coherent part of the electromagnetic field, in the same fashion as the dispersive coupling contribution, whereas the latter is related to the waveguide-cavity scattering mechanism and is central to our discussion. In fact, for largely off-resonant excitation, which is the relevant case in sideband-resolved optomechanical systems, we have $\bar{a} \approx  \frac{ \sqrt{\bar{\kappa}_e}\bar{\alpha}_\text{in}}{i\bar\Delta}$. This is the enhancement factor for dispersive optomechanical coupling. In the case of waveguide-cavity scattering, the expression playing this same role is $ \frac{\bar{\alpha}_\text{in}}{2\sqrt{\kappa_e}}$, which is a factor $\approx \Delta/(2\kappa_e)$ larger than its dispersive counterpart. In our experiment, $\Delta/(2\kappa_e)\approx 20$, justifying the large contribution from dissipative optomechanics at large detunings. Near the resonance, however, this term decreases, and so does the dissipative contribution.

 The photocurrent generated at our detector is a function of the field reflected by our resonator. This is obtained through input-output relations:

 \begin{equation}
     \alpha_\text{out}(t) =  \bar{\alpha}_\text{in} + \sqrt{\bar{\kappa}_e} \left(\bar{a}+\delta a(t)\right)+\frac{G_{\kappa_e}\bar{a}}{2\sqrt{\bar{\kappa}_e}} \delta x (t).
     \label{eq:outp}
 \end{equation}
 The output field $\alpha_\text{out}(t)$ can also be linearized as $\alpha_\text{out}(t) = \bar{\alpha}_\text{out} + \delta \alpha_\text{out}(t)$, and fluctuating terms can be matched between the right-hand and left-hand sides of Eq.~\ref{eq:outp}. The photocurrent $I(t)$ is proportional to $|\alpha_\text{out}(t)|^2$, whose fluctuations are given by $\delta I(t) = \bar{\alpha}_\text{out} \delta \alpha^*_\text{out}(t)+\bar{\alpha}^*_\text{out} \delta \alpha_\text{out}(t)$. Using a spectrum analyzer, one measures the power spectral density of $\delta I(t)$, $S_{II}(\omega) = \int d \tau e^{i \omega \tau}\langle \delta I(\tau) \delta I(0)\rangle$. Finding this quantity requires moving into a frequency domain description in Eq.~\ref{eq:dyn}, where the optical field $\delta a(\omega)$ can be written solely in terms of $\delta x(\omega)$. We find 

\begin{equation}
    S_{II}(\omega) = \frac{16 |\bar{\alpha}_\text{in}|^4 |\left(-\bar{\kappa} _e+\bar\kappa -i \omega \right)\left(4 \bar{\Delta} ^2 G_{\kappa _e}-8 \bar{\Delta}  \bar{\kappa}
   _e G_{\omega }+\left(\bar{\kappa} -2 \bar{\kappa}_e\right) (\bar\kappa -2 i \omega ) G_{\kappa _e}\right)|^2 }{\left(4 \bar{\Delta} ^2+\bar{\kappa} ^2\right)^2 \left(8 \bar{\Delta} ^2 \left(\bar{\kappa} ^2-4 \omega
   ^2\right)+16 \bar{\Delta} ^4+\left(\bar{\kappa} ^2+4 \omega ^2\right)^2\right)} S_{xx}(\omega),
\label{eq:Sii}
\end{equation}
where $S_{xx}$ is the mechanical power spectral density. Assuming a weak optomechanical coupling, $S_{xx}$ is related to the bare mechanical response, in the absence of optomechanical backaction, and therefore is independent of $\bar{\Delta}$. As a consequence, the prefactor of $S_{xx}$ in Eq.~\ref{eq:Sii} can be used as a model for optomechanical transduction. Here, we omitted a prefactor related to the particularities of the detection system used, such as signal losses, photodetector gain, and efficiency. 

\section{Dynamical backaction in a dissipative optomechanical system}
The Hamiltonian of a generalized optomechanical system in the presence of both dissipative and dispersive couplings is given by: 
$\hat{H} = \hbar\omega_c \hat{a}^\dagger\hat{a} +\hbar\Omega \hat{b}^\dagger\hat{b} + \sum_{q} \hbar \omega_{q,i} \hat{c}_{q,i}^\dagger \hat{c}_{q,i} +\sum_{q} \hbar \omega_{q,e} \hat{c}_{q,e}^\dagger \hat{c}_{q,e}+ \hat{H}_\kappa + \hat{H}_{\Gamma} + \hat{H}_\mathrm{OM}$. The first two terms in $\hat{H}$ describe the bare optical and mechanical oscillators. $\hat{H}_\kappa$ and  $\hat{H}_{\Gamma}$ describe the (unperturbed) damping of the optics ($\kappa$) and mechanics ($\Gamma$), respectively, which carry no dependency on $\hat{x}$. In particular, optical losses have both intrinsic and extrinsic contributions, related to the third and fourth terms describing the respective bosonic baths with subscripts ``$i$" and ``$e$". Assuming dispersive and extrinsic dissipative optomechanical interactions, the interaction Hamiltonian $\hat{H}_\mathrm{OM}$ has the form:
\begin{equation}
    \hat{H}_\text{OM} = -\left[\hbar G_\omega \hat{a}^{\dagger} \hat{a}+ i\sqrt{\frac{\kappa_e}{2 \pi \rho_e}} \frac{\hbar G_{\kappa_e}}{2\kappa_e} \sum_{q}\left(\hat{a}^{\dagger} \hat{c}_{q,e}-\hat{c}_{q,e}^{\dagger} \hat{a}\right)\right]\hat x.
    \label{Hamiltonian}
\end{equation}
 Here, $\hat x$ is the mechanical position operator and $\rho_e$ denotes the density of states of the waveguide. In the Markovian limit ($\omega_c \gg \kappa $), we may treat $\rho_e$ as a constant. We assume only the coupling to the extrinsic bath modes to be affected by the mechanical displacement, ignoring the so-called intrinsic dissipative coupling. Tracing out the optical baths by means of the input-output formalism, the linearized optical force operator is $\hat F = - \frac{\partial \hat H_\text{OM}}{\partial \hat x}$~\cite{Elste2009QuantumNanomechanics, Weiss2013Strong-couplingSystems}

\begin{equation}
\begin{aligned}
\hat{F} x_{\text{zpf}} &=\hbar g_\omega \left(\bar{a}^{*} \delta\hat{a}+\bar{a} \delta\hat{a}^{\dagger}\right) \\
&+i \frac{\hbar g_{\kappa_e}}{2\kappa_e} \sqrt{\kappa_{\text {e}}}\left[\bar{a}^{*} \delta \hat{\alpha}_{\text {in}}^{\text{e}}-\left( \delta \hat{\alpha}_{\text {in}}^{\text{e}}\right)^{\dagger} \bar{a}\right] 
+i\frac{\hbar g_{\kappa_e}}{2\kappa_e}\sqrt{\kappa_e}\left[\bar{\alpha}_\text{in}\delta \hat{a}^{\dagger}-\bar{\alpha}_\text{in}^{*}\delta\hat{a}\right].
\label{eq:force_op}
\end{aligned}
\end{equation}
Here, we linearized the optical dynamics around coherent ($\bar a = \langle\hat a\rangle$) and fluctuation ($\delta \hat a$) amplitudes, in the form of $\hat a = \bar{a}+  \delta \hat a$. The dissipation-fluctuation theorem~\cite{Kubo1966TheTheorem} shows that the coupling to the bath modes will effectively add noise to the optical field (modeled as a fluctuating source term). In our system, this noise is intrinsic ($\delta \hat{\alpha}_{\mathrm{in}}^{\text{i}}$) or extrinsic ($\delta \hat{\alpha}_{\mathrm{in}}^{\text{e}}$) in nature. Since the coupling to the extrinsic bath is modulated by the mechanics, $\delta \hat{\alpha}_{\mathrm{in}}^{e}$ also appears in the force operator. 

The first line in Eq.~\ref{eq:force_op} describes the force arising from the dispersive coupling, while the second is related to the extrinsic dissipative coupling. The last term is particularly interesting since it gives rise to a force directly dependent on the coherent drive amplitude through the extrinsic channel. For telecom optical photons, we may assume a zero temperature optical bath, with noise correlations: $\left\langle\delta \hat{\alpha}_{\mathrm{in}}^{J}(\omega) (\delta \hat{\alpha}_{\mathrm{in}}^{J^\prime}\left(\omega^{\prime})\right)^{\dagger}\right\rangle=2 \pi \delta\left(\omega+\omega^{\prime}\right)\delta_{J,J^{\prime}}$, $J= \text{e},\text{i}$, meaning the intrinsic and extrinsic noises are uncorrelated.
In the weak coupling regime ($C \ll \kappa/\Gamma$), one may compute the backaction cooling/amplification of the mechanical mode from the unperturbed force's noise spectrum $S_\mathrm{FF}(\Omega)$\cite{Elste2009QuantumNanomechanics, Weiss2013Strong-couplingSystems, Primo2020Quasinormal-ModeOptomechanics} and the Fermi's Golden Rule. In the present case, we have $\delta\Gamma=\frac{x_\mathrm{zpf}^2}{\hbar^2} \left[
 S_\mathrm{FF}(\Omega)  - S_\mathrm{FF}(-\Omega) \right]$. Plugging in the expression for $S_\mathrm{FF}$, we have

\begin{equation}
\delta\Gamma  = -\frac{4 n_c \Omega}{\bar{\kappa}_e^2} \frac{\left\{2 g_{\omega } g_{\kappa _e} \bar{\kappa}_e \left[\bar{\kappa} _e \left(-4 \bar{\Delta} ^2+\bar{\kappa}^2+4
   \Omega ^2\right)-8 \bar{\Delta}^2 \bar{\kappa} \right]+\bar{\Delta}  g_{\kappa _e}^2 \left[4 \bar{\Delta}^2 \bar{\kappa}
   +\bar{\kappa}_e \left(4 \bar{\Delta}^2-3\bar{\kappa}^2-4 \Omega ^2\right)+\bar{\kappa}^3\right]+16 \bar{\Delta} \bar{\kappa}_e^2 \bar{\kappa} 
   g_{\omega }^2\right\}}{16 \bar{\Delta}^4+8 \bar{\Delta} ^2 \left(\bar{\kappa} ^2-4 \Omega ^2\right)+\left(\bar{\kappa} ^2+4 \Omega
   ^2\right)^2},
   \label{eq:deltaGamma}
\end{equation}
where we clearly see an interplay between dispersive and extrinsic dissipative couplings. In our case, $g_{\kappa_e} \ll g_\omega$, and the main dissipative contributions to dynamical backaction come through crossed terms of the form $g_\omega g_{\kappa_e}$. Taking $\bar{\Delta} = \pm \Omega \gg \kappa$ we recover the results in the main text. 

In the main text, we use Eq.~\ref{eq:deltaGamma} to fit the values of $\delta\Gamma$ extracted from our experiment. A taper-waveguide coupling efficiency calibration, along with the mechanical and optical spectra limit our fitting parameters to $g_\omega$ and $g_{\kappa_e}$.

\section{Optomechanically induced transparency}

In this section, we derive the equations for the dressed optical susceptibility due to the optomechanically induced transparency/absorption phenomena. We start with the important remark that in coherent spectroscopy the input laser is phase modulated, that is, one needs the substitution $\bar{\alpha}_\text{in} \to \bar{\alpha}_\text{in}e^{-i \phi_0 \sin(\Omega_\text{mod}t)}$ in Eq.~\ref{eq:rate_a}. For weak modulations, $\phi_0 \ll 1$, we may approximate $\bar{\alpha}_\text{in}e^{-i \phi_0 \sin(\Omega_\text{mod}t)} \approx \bar{\alpha}_\text{in} + \frac{\phi_0 \bar{\alpha}_\text{in}}{2} e^{-i \Omega_\text{mod}t}-\frac{\phi_0 \bar{\alpha}_\text{in}}{2} e^{i \Omega_\text{mod}t}$, and the system is effectively driven by a strong pump tone, at frequency $\omega_l$ and two probes at $\omega_l \pm \Omega_\text{mod}$. 

The optical response to the phase-modulated input is modeled doing $a \to \bar{a} + a_+ e^{-i \Omega_\text{mod}t} + a_- e^{i \Omega_\text{mod}t}$. In order to simplify our discussion, we assume the pump laser is detuned several linewidths away from our optical resonance and $\Omega_\text{mod}\gg \kappa$, in accordance with experimental conditions. In this case, the optical susceptibility acts as a bandpass filter, effectively filtering one of the sidebands of our probe tones. Assuming a red-detuned laser (the OMIT configuration), the phonon-creation process is suppressed and only $a_+$ is appreciable. Eq.~\ref{eq:dyn} is rewritten as:

\begin{equation}
    -i \Omega_\text{mod} a_+ =  i (\bar \Delta  a_+ + g_\omega \bar{a} b_+ )  - \frac{\bar\kappa a_+ + g_{\kappa_e}\bar{a} b_+}{2} - \frac{g_{\kappa_e} b_+}{2\sqrt{\bar{\kappa}_e}}  \bar{\alpha}_\text{in} - \sqrt{\bar{\kappa}_e} \frac{\phi_0}{2}\bar{\alpha}_\text{in},
    \label{eq:rateOMIT}
\end{equation}
where $b_+$ is the amplitude of the phononic field oscillating at $\Omega_\text{mod}$. This quantity is obtained by solving the (classical) Langevin equation for $b$ (obtained through the Hamiltonian of the system) with the substitution $b \to \bar{b} + b_+ e^{-i\Omega_\text{mod}t}+ b_- e^{i\Omega_\text{mod}t}$.

The beating between the pump and probe tones will give rise to coherent mechanical oscillations, mediated by the optical force. In this classical picture, and under the same approximations above, Eq.~\ref{eq:force_op} becomes

\begin{equation}
\begin{aligned}
   F x_{\text{zpf}} &=\hbar g_\omega \bar{a}^{*} a_+ e^{-i \Omega_\text{mod}t}\\
   &+ i \frac{\hbar g_{\kappa_e}}{2\sqrt{\bar{\kappa}_e}}\left[\bar{\alpha}_{\text{in}}\bar{a}^{*}  \left(\frac{\phi_0}{2} e^{-i \Omega_\text{mod}t} -\frac{\phi_0}{2} e^{i \Omega_\text{mod}t}\right)-\left(\frac{\phi_0}{2} e^{-i \Omega_\text{mod}t}-\frac{\phi_0}{2} e^{i \Omega_\text{mod}t}\right)^{*} \bar{\alpha}^{*}_{\text{in}}\bar{a}\right] 
-i\frac{\hbar g_{\kappa_e}}{2\sqrt{\bar{\kappa}_e}}\bar{\alpha}_\text{in}^{*}a_+ e^{-i \Omega_\text{mod}t}.
\label{eq:Fx}
\end{aligned}
\end{equation}
In our system, the second term, which is rooted in the beating between the fluctuations (or oscillations) in the input and cavity fields is necessarily smaller than the other two. This is the case since it arises from the dissipative optomechanical coupling and scales with $\bar{a}$. For instance, since $g_{\kappa_e}\ll g_\omega$, the dispersive term (which also scales with $\bar{a}$) is obviously dominant over it. Furthermore, the last term, which is also dissipative in nature, scales with the photon flux in the waveguide, drastically increasing its importance relative to the second term. Joining all this information and writing the equation for $b_+$, we have

\begin{equation}
    -i\Omega_\text{mod} b_+ = -\left(i \Omega +\frac{\Gamma}{2}\right)b_+ + i g_\omega \bar{a}^{*} a_+ +\frac{ g_{\kappa_e}}{2\sqrt{\bar{\kappa}_e}}\bar{\alpha}_\text{in}^{*}a_+,
    \label{eq:b-dyn}
\end{equation}
which can be solved for $b_+$ in terms of $a_+$. Substituting this result back in Eq.~\ref{eq:rateOMIT} yields a dressed susceptibility for $a_+$, i.e.

\begin{equation}
    a_+ = -\frac{\sqrt{\bar{\kappa}_e}\bar{\alpha}_\text{in}\phi_0}{-2i\left( \bar{\Delta} + \Omega_\text{mod}\right) + \bar{\kappa} - \frac{n_c \left[( \bar{\kappa} +2 i \bar{\Delta} ) g_{\kappa _e}-4 i \bar{\kappa}_e g_{\omega
   }\right] \left[\left(2 i \bar{\Delta} +2 \bar{\kappa}_e-\bar{\kappa} \right) g_{\kappa _e}-4 i
   \bar{\kappa}_e g_{\omega }\right]}{4 \bar{\kappa}_e^2 \left(\Gamma -2 i \Omega
   _{\text{mod}}+2 i \Omega \right)}}.
   \label{eq:a+}
\end{equation}
Here, we see that the cavity response is modified by the optomechanical coupling.  This is most relevant at $\bar{\Delta} = -\Omega_\text{mod} = -\Omega$. In this case, and assuming a system well into the sideband-resolved regime, we have

\begin{equation}
    a_+ \approx - \frac{\sqrt{\bar{\kappa}_e}\bar{\alpha}_\text{in}\phi_0}{\bar{\kappa} + \frac{n_c \left(2 \bar{\kappa} _e g_{\omega }+\Omega  g_{\kappa
   _e}\right){}^2}{\Gamma  \bar{\kappa} _e^2}},
   \label{eq:a+simpl}
\end{equation}
which shows an effective enhancement in the cavity losses, or, equivalently a transparency window~\cite{Safavi-Naeini2011ElectromagneticallyOptomechanics,Weis2010OptomechanicallyTransparency}. This is equivalent to what has been verified for dispersive optomechanical systems, with the addition of the dissipative contribution. We notice that, in agreement with the discussion in the main text, the transparency window is favored by $g_\omega$ and $g_{\kappa_e}$ with equal signs. This is also the case of backaction cooling under red-detuned excitation, as mentioned in the main text. 

The generalization of the expressions above accounting for multiple, independent, mechanical modes (as is the case in our experiment) is done by simply adding extra optomechanical contributions to the optical response in the denominators of Eqs.~\ref{eq:a+} and  \ref{eq:a+simpl}, with modified mechanical linewidths, frequencies, and optomechanical couplings.

With this result, we are finally able to describe the scattering parameter $|S_{21}(\Omega_\text{mod})|$ of the main text. This is done by using input-output relations, accounting for the phase-modulated laser. The outgoing field amplitude, $\alpha_\text{out}$ reads

\begin{equation}
    \alpha_\text{out} =  \alpha_\text{in} + \sqrt{\kappa_e} a + \frac{G_{\kappa_e} a}{2 \sqrt{\kappa_e}}x,
\end{equation}
where the last term arises from the dissipative optomechanical coupling. In the linearized regime, this term is of little relevance, since it combines the intra-cavity field $\bar{a}$ and the dissipative coupling, which is small. Furthermore, it is not relevant for the modification of the cavity's susceptibility: it is generated by photons scattered from the cavity into the waveguide and therefore carries no information from the fluctuations in the cavity field. In that spirit, we neglect it in our analysis.

The outgoing field amplitude will have components at $\omega_l$ and $\omega_l \pm \Omega_\text{mod}$, i.e. $\alpha_\text{out} = \bar{\alpha}_\text{out} + \alpha_{\text{out},+} e^{-i \Omega_\text{mod}t} + \alpha_{\text{out},-} e^{i \Omega_\text{mod}t}$. Combining this with the substitutions for the input and cavity fields, we have
\begin{align}
    &\bar{\alpha}_{\text{out}} =  \bar{\alpha}_\text{in} + \sqrt{\kappa_e} \bar{a} =  \bar{\alpha}_\text{in}r(\bar\Delta),\\
    &\alpha_{\text{out},+} =  \frac{\phi_0 \bar{\alpha}_\text{in}}{2} + \sqrt{\kappa_e} a_+ =  \frac{\phi_0 \bar{\alpha}_\text{in}}{2} r_\text{OM}(\bar\Delta + \Omega_\text{mod}),\\
    &\alpha_{\text{out},-} =  -\frac{\phi_0 \bar{\alpha}_\text{in}}{2} + \sqrt{\kappa_e} a_- = -\frac{\phi_0 \bar{\alpha}_\text{in}}{2} r_\text{OM}(\bar\Delta - \Omega_\text{mod}),
\end{align}
where we defined $r$ and $r_\text{OM}$ as the ``bare" and dressed cavity reflectivities, respectively. Notice that for a red-detuned laser with $\bar\Delta =  - \Omega$ and modulation frequency $\Omega_\text{mod} =  \Omega$, the on-resonance reflectivity is probed through $r_\text{OM}(\bar\Delta + \Omega_\text{mod})$, which is directly related to $a_+$. This is not the case for $r_\text{OM}(\bar\Delta - \Omega_\text{mod})$, which gives the response of the cavity two mechanical frequencies away from resonance, where the optomechanical contribution to the optical susceptibility is negligible. In fact, in this case, a reasonable approximation is $r_\text{OM}(-2\Omega) \approx r(-2\Omega)$.  

The scattering matrix element $|S_{21}(\Omega_\text{mod})|$ is directly proportional to the output field amplitude at frequency $\Omega_\text{mod}$~\cite{Shomroni2019OpticalOscillator}. This is given by

\begin{equation}
     |S_{21}(\Omega_\text{mod})| = \frac{\phi_0 |\bar{\alpha}_\text{in}|^2}{2}\left|r(\bar\Delta) r_\text{OM}^*(\bar\Delta-\Omega_\text{mod})-r^*(\bar\Delta) r_\text{OM}(\bar\Delta+\Omega_\text{mod})\right|,
     \label{eq:S21}
\end{equation}
where we omitted a scale factor. In our fittings throughout the text, an additional phase factor is introduced in order to handle any difference in the dispersion of the sidebands. This is done by making $r_\text{OM}(\bar\Delta+\Omega_\text{mod}) \to r_\text{OM}(\bar\Delta+\Omega_\text{mod}) e^{i \psi}$. Our fitting parameters are chosen to be $\bar\Delta$, $\psi$, and $g_\omega$, with all the other parameters extracted from independent measurements shown in Figs.~2-4 in the main text.

\section{Analysis for Acoustic Mode 2}
The results in Section~\ref{S1} can be readily extended to the more realistic scenario in which each optical resonator is coupled to its own acoustic mode. Consider the system of coupled cavities in Fig.~\textbf{1b} of the main text, where each optical resonator displays different resonance frequencies $\omega_i$, $i = 1,2$, and is coupled to independent acoustic modes with mechanical displacement $x_i$ through a dispersive coupling $G_{\omega_i}$. The generalization of the matrix $\mathbf{M}$ for this particular system is given by
\begin{equation}
\mathbf{M} = 
    \begin{bmatrix}
        -i \left(\omega_1- G_{\omega_1} x_1\right) - \frac{(\kappa_e+\kappa_i)}{2} & iJ\\
        iJ & -i\left( \omega_2 - G_{\omega_2}x_2 \right) - \frac{\kappa_i}{2}
    \end{bmatrix}.
\end{equation}
 Once more, we are interested in the case where the frequencies of the bare optical resonators are degenerate, i.e. $\omega_2=\omega_1$. Taylor expanding $\kappa_{\mp}$ and $\omega_{\mp}$ up to first order in the $x_i$, we obtain:
\begin{equation}
\begin{split}    
    \kappa_{\pm} &= \kappa_i + \kappa_e\left(\frac{1}{2} \mp \frac{G_{\omega_1}}{4\sqrt{J^2 - (\kappa_e/4)^2}}x_1 \pm \frac{G_{\omega_2}}{4\sqrt{J^2 - (\kappa_e/4)^2}}x_2\right),\\
    \omega_{\pm} &= \omega_1 \pm 4\sqrt{J^2 - (\kappa_e/4)^2}-\frac{1}{2}\left(G_{\omega_{1}}x_1 + G_{\omega_{2}}x_2\right).
\end{split}
\end{equation}
One can then define a dissipative (dispersive) coupling $G_{\kappa_{e_{\pm}}}^{(i)}$($G_{\omega_{\pm}}^{(i)}$) associated to acoustic mode $i$:
\begin{align}
G_{\kappa_{e_\pm}}^{(i)} &= \frac{d\kappa_\pm}{dx_{i}} = \begin{cases}
\mp\frac{G_{\omega_i} \kappa_e}{4\sqrt{J^2 - (\kappa_e/4)^2}}, & i\text{ = 1},\\
\pm\frac{G_{\omega_i} \kappa_e}{4\sqrt{J^2 - (\kappa_e/4)^2}}, & i\text{ = 2},\end{cases}\\
G_{\omega_\pm}^{(i)} &=- \frac{d\omega_\pm}{dx_{i}} = \frac{G_{\omega_i}}{2}.
\label{eq:Gkmode2}
\end{align}
In addition to recovering Eq.~\ref{eq:gk_theory} in the $G_{\omega_2} \to 0$ limit, Eq.~\ref{eq:Gkmode2} also reveals that $G^{(1)}_{\kappa_e{_{\pm}}} = -G^{(2)}_{\kappa_e{_{\pm}}}$, meaning the sign of the dissipative coupling is flipped between different acoustic modes. This asymmetry between the behaviors of mechanical modes $1$ and $2$ is generated due to the fact that only one of the optical cavities is coupled to the waveguide. Furthermore, the dispersive coupling maintains its sign irrespective of the mechanical modes,  $G^{(1)}_{\omega_{\pm}} = G^{(2)}_{\omega_{\pm}}$, implicating that optomechanical interactions dependent on the interplay between dispersive and dissipative coupling will display opposing behaviors between the different mechanical modes. In our experiment, acoustic modes $1$ and $2$ correspond to the independent mechanical breathing modes of the nanobeams. A frequency mismatch between them, due to fabrication fluctuations, enables their individual analysis of the impacts of the mechanically mediated waveguide-cavity interaction.

\section{Nearly cold-cavity transduction}
Due to the high Q-factors and small modal volumes of our silicon devices, powers as low as $\SI{1}{\micro\watt}$ trigger nonlinear optical effects such as thermo-optical dispersion, two-photon absorption, and (consequently) free-carrier absorption and dispersion~\cite{Primo2021AccurateOptomechanics, Barclay2005NonlinearTaper}. This is manifested in the optical spectra of our device, which is shown to be bistable in Fig.~3\textbf{b} of the main text. 

\begin{figure*}[h]
\centering
\includegraphics[width = 16cm]{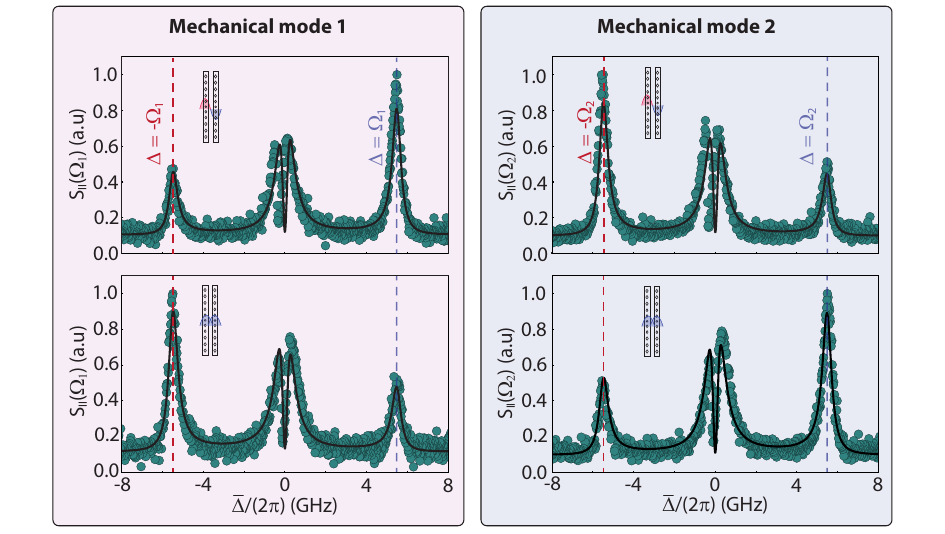}
\caption{\small{Optomechanical transduction for mechanical modes $1$ (left) and $2$ (right), under excitation of the antisymmetric (top) and symmetric (bottom) optical modes.}}
\label{fig:1}
\end{figure*}

In Fig.~\ref{fig:1} we show the transduction spectra of mechanical modes $1$ and $2$ outside the optical bistable regime. We readily verify that the same asymmetries between red and blue-detunings is maintained for all combinations of optical and acoustic modes, in accordance with previously shown results. Furthermore, the black curves are the same theoretical estimates used before which were based on the values of dissipative and dispersive couplings extracted from backaction/OMIT experiments. The agreement between theory and experiment is poorer than for higher-power data. We attribute this to the low signal-to-noise ratio of the mechanical spectra for low optical powers and to a slight thermo-refractive shift of the resonance under near-resonant excitation.

\section{Independent control of degenerate acoustic modes}

The interplay between dissipative and dispersive couplings allows for the augmentation or suppression of backaction-induced heating or cooling effects, depending on the relative sign between $g_\omega$ and $g_{\kappa_{e}}$, as demonstrated in the main text. For sufficiently high values of $g_{\kappa_e}$, a near-perfect cancellation of dynamical backaction can be achieved for a given optical/mechanical mode pair, enabling the independent control of virtually degenerate acoustic modes. This regime is obtained when the dissipative-to-dispersive coupling ratio is given by

\begin{equation}
    \biggr|\frac{g_{\kappa_e}}{g_{\omega}}\biggr| = \frac{8\Omega_m^2 - \bar{\kappa}\bar{\kappa}_e}{\bar{\kappa}(\bar{\kappa}-3\bar{\kappa}_e)\Omega_m +4\Omega_m^3} \implies \frac{\partial \delta \Gamma}{\partial g_{\kappa_e}}\biggr|_{\pm \Omega_m}  = 0,
\end{equation}
as can be derived from Eq.~\ref{eq:deltaGamma}. This is the point where variations on the mechanical linewidth are insensitive to an increase/decrease in the dissipative coupling and therefore constitute a minimum in $|\delta\Gamma|$ (given all other parameters are kept fixed).

\begin{figure*}[h]
\centering
\includegraphics[width = 6.97cm]{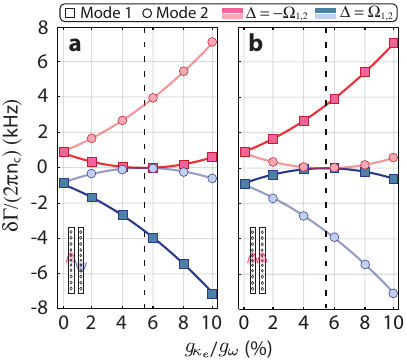}
\caption{\small{\textbf{a} (\textbf{b}) Theoretical optically-induced modifications to the mechanical linewidth of mechanical modes 1 and 2 for the differential (common) optical modes scaled by the intracavity photon number, as a function of dissipative-to-dispersive optomechanical coupling ratio. Dashed line demarcates the threshold of near cancellation of either the backaction heating or cooling in a given optical/mechanical mode pair. Optical and mechanical parameters: $\Gamma_\text{m}/2\pi = \SI{2.6}{MHz}$, $\Omega_{1,2}/2\pi = \SI{5.5}{GHz}$, $\bar{\kappa}/2\pi = \SI{550}{MHz}$, $\bar{\kappa}_e/2\pi = \SI{150}{MHz}$, $g_{\omega}/2\pi = \SI{350}{kHz}$.}}
\label{fig:2}
\end{figure*}

Fig.~\ref{fig:2} illustrates the theoretical optically-induced heating and cooling phenomena observed in mechanical modes 1 and 2 under both blue and red excitations. Remarkably, by considering optical and mechanical modes' properties akin to those outlined in the main text, a notable suppression of backaction effects can be realized at a critical ratio of $|g_{\kappa_e}/g_{\omega}| \approx 5.4\%$.


As an example, Fig.~\ref{fig:2}\textbf{a} demonstrates that when driving the differential optical mode at $\Delta = -\Omega_{1,2}$ one achieves enhanced cooling in acoustic mode 2 while strongly suppressing it in acoustic mode 1 at a critical ratio of $|g_{\kappa_e}/g_{\omega}| \approx 5.4\%$. Repeating this analysis for $\Delta = \Omega_{1,2}$ (still considering the differential optical mode) signals a strong reduction of backaction heating in mode 2. As expected, opposite trends are showcased in Fig.~\ref{fig:2}\textbf{b} for the common optical mode. This decoupling between the responses of nearly degenerate acoustic modes suggests that for a given optical detuning it is possible to individually address and modify their states. This feature highlights one promising application in the field of quantum optomechanics, where information could be written and retrieved at will from specific mechanical modes, which could be therein used as versatile quantum memories~\cite{Wallucks2020AWavelengths} and/or building blocks for microwave-to-optical transducers~\cite{Jiang2022OpticallyPhotons}.
\color{black}

\textit{Note.} -- Data, FEM and script files for generating each figure will be available at the \textit{Zenodo} repository.~\cite{zenodo_data}.

\section*{Supplementary References}

%